\documentclass[aps,pra,preprint, amsmath,amssymp,showpacs]{revtex4-1}

\usepackage[T1]{fontenc}
\usepackage[latin1]{inputenc}
\usepackage{rotating}
\usepackage{bm}        
\usepackage{amssymb}   
\usepackage{amsmath} 
\usepackage{subfig}

\usepackage{physics}

\usepackage{caption}

\usepackage{bm}
\usepackage{CJK}
\def\jcp#1#2#3{J.~Chem.~Phys.~{\bf #1},\ #2\ (#3)}

\def\pra#1#2#3{Phys.~Rev.~A~{\bf #1},\ #2\ (#3)}
\def\prl#1#2#3{Phys.~Rev.~Lett.~{\bf #1},\ #2\ (#3)}

\def\k1{k_1}
\def\k2{k_2}
\def\q1{q_1}
\def\q2{q_2}

\def\({\left (}
\def\){\right )}
\def\[{\left [}
\def\]{\right ]}

\newcommand{\beq}{\begin{equation}}
\newcommand{\eeq}{\end{equation}}

\begin{document}
\date{\today}
\flushbottom \draft
\title{Long-lived quantum coherences in a V-type system strongly driven by a thermal environment} 

\author{Suyesh Koyu and Timur V. Tscherbul}
\affiliation{Department of Physics, University of Nevada, Reno, NV 89557, USA}\email[]{ttscherbul@unr.edu}

\begin{abstract}

We explore the coherent dynamics of a three-level V-system interacting with a thermal bath in the  regime where thermal excitation occurs much faster than spontaneous decay. We present analytic solutions of the Bloch-Redfield quantum master equations, which show that strong incoherent pumping can generate long-lived quantum coherences among the excited states of the V-system in the overdamped regime defined by the condition $\Delta/(\bar{n}\gamma)<f(p)$, where $\Delta$ is the excited-state level splitting, $\gamma$ is the spontaneous decay rate, $\bar{n}\gg 1$ is the effective photon occupation number proportional to the pumping intensity, and $f(p)$ is a universal function of the transition dipole alignment parameter $p$. In the limit of nearly parallel transition dipoles ($p\to 1$) the coherence  lifetime $\tau_c = 1.34 (\bar{n}/\gamma) (\Delta/\gamma)^{-2}$ scales linearly with $\bar{n}$ and is enhanced by the factor  $0.67 \bar{n}$  with respect to the weak-pumping limit [{Phys.~Rev.~Lett.} {\bf 113}, 113601 (2014); \jcp{144}{244108}{2016}]. 
We also establish the existence of long-lived quasistationary states, which occur  in the overdamped regime and affect the process of thermalization of the V-system with the bath, slowing down the approach to thermal equilibrium.
In the case of nonparallel transition dipole  moments ($p<1$), no quasistationary states are formed and the coherence lifetime decreases sharply.   Our results reveal new regimes of long-lived quantum coherent dynamics, which could be observed  in thermally driven atomic and molecular systems.
\end{abstract}


\maketitle
\clearpage
\newpage

\section{introduction}

Relaxation and loss of coherence in multilevel quantum systems caused by their interaction with a thermal environment is a subject of paramount importance in many areas of physics including quantum optics \cite{Schlosshauer,BPbook,ScullyBook}, quantum sensing \cite{Qsensors}, and quantum  information processing \cite{Ladd}. 
While interaction with the environment is generally believed to destroy any quantum coherence initially present in the system \cite{Schlosshauer}, recent theoretical studies have challenged this point of view    suggesting a number of mechanisms for the generation of quantum (Fano) coherences in multilevel systems driven by thermal noise \cite{Scully92,HegerfeldtPlenio93,AgarwalMenon01,Keitel05, Scully06,Ou08,Scully11,Scully13,prl14,Amr16,Amr16slow}.  These mechanisms have attracted attention due to their predicted ability to enhance the efficiency of quantum heat engines \cite{Scully11,Scully13} and as potential sources of non-trivial quantum effects in photosynthetic light-harvesting \cite{prl14,Amr16,Amr16slow}.

The noise-induced Fano coherences can be understood as arising from quantum interference of the different incoherent excitation pathways originating from the same initial state \cite{Scully06,Scully92}. The mathematical description of the interference effects requires the use of non-secular Bloch-Redfield (BR) theory, in which populations and coherences are treated on the same footing,  leading to more complex dynamics than predicted by the secular rate equations  \cite{Amr16,prl14,AgarwalTract}. Such noise-induced coherent dynamics are responsible for a number of remarkable effects such as vacuum-induced coherence \cite{Evers13}, enhanced efficiency of quantum heat engines \cite{Scully11,Scully13}, and long-lived quasistationary states \cite{prl14}.  Note that the secular approximation  cannot be justified in systems with nearly degenerate energy levels, where the system evolution time  can be much longer than the timescale of interest \cite{prl14}.

The three-level V-system comprising a single ground state coupled  by the system-bath interaction to a pair of excited states (see Fig. 1)  serves as a minimal model of a multilevel quantum system exhibiting non-trivial Fano coherence dynamics. This system has been extensively studied  in the weak-pumping limit (relevant for photosynthetic light-harvesting) where incoherent excitation occurs much more slowly than  spontaneous emission. In this limit, the coherent dynamics of the V-system is determined by  the ratio $\zeta=\frac{1}{2}(\gamma_a+\gamma_b)/\Delta_p$, where $\Delta_p=\sqrt{\Delta^2+(1-p^2)\gamma_a\gamma_b}$ is the renormalized excited-level splitting, $\gamma_a$ and $\gamma_b$   are the spontaneous decay rates, and $p$ is the angle between the transition dipole moments of the $g\to a$ and $g\to b$ transitions  (see Fig. 1) \cite{Amr16}.  The two-photon coherences between the excited states of the V-system exhibit damped oscillations in the regime where the excited levels are widely spaced ($\zeta\ll 1$). In the opposite regime  of small level spacing ($\zeta\gg 1$), the coherences evolve monotonously and can survive for an arbitrarily long time $\tau_c=2\sqrt{\gamma_a\gamma_b}/\Delta_p^2$ \cite{prl14,Amr16}.



While the weak pumping regime of  noise-induced coherent dynamics is well understood \cite{Scully06,prl14,Amr16,Amr16slow}, much less is known about the opposite limit where incoherent excitation occurs much faster than spontaneous emission. The strong pumping  regime is central to the theory of quantum heat engines, where quantum coherence has been  predicted to enhance the engine's efficiency  \cite{Scully11,Scully13}. Accordingly, the generation and steady-state properties of  quantum coherences in this regime have been studied in closed-cycle quantum heat engine models \cite{Scully11} and in the degenerate $\Lambda$-system \cite{Ou08}. However, these studies did not explore the time dynamics of the coherences as a function of the system's excited-state splitting and radiative decay rates. In addition, the quantum heat engine studies \cite{Scully11,Scully13}  considered a more complex case of a 5-level system interacting with two baths, where the coherences emerge as a result of non-equilibrium transport dynamics involving both of the baths.  This leaves open the  question of whether  strong incoherent driving  can generate coherences in multilevel quantum systems interacting with a single thermal bath.


Here, we address this question by presenting a theoretical analysis of the quantum dynamics of a  V-system strongly driven by a thermal bath. We derive closed-form analytic  solutions of the Bloch-Redfield (BR)  quantum master equations, which show that (1) quantum coherences can be generated by strong incoherent driving provided that the transition dipole moments of the V-system are nearly perfectly aligned, and (2) the coherence lifetime scales linearly with the pumping intensity $\bar{n}$ and quadratically with the inverse excited-level spacing $\gamma/\Delta$. These results suggest the possibility of observing  long-lived  coherence dynamics in strongly driven Rydberg atoms and polyatomic molecules.  


This paper is structured as follows. In Sec. II we present the theoretical formalism based on the BR master equations and outline the procedure of their analytical solution. The dynamical regimes of the strongly driven V-system are classified in Sec. IIB. Sections IIC and III present analytical expressions for the coherence lifetimes and for the time dynamics of the populations and coherences. Section IV summarizes the main fundings of this work and outlines an experimental scenario for observing the noise-induced coherences.

\section{Theory}


\subsection{Bloch-Redfield equations and their general solution}

Consider a three-level V-system weakly coupled to a thermal environment (see Fig. 1). The system  resides in the ground state $|a\rangle$ ({\it i.e.} $\rho_{aa}(0)=1$) before the system-environment coupling  is  suddenly turned on at $t=0$, leading to the population transfer to the excited states $|b\rangle$ and $|c\rangle$. To describe the time evolution of the system, we use a quantum master equation approach based on the Liouville-von Neumann equation for the density operator of the  system+bath complex  \cite{BPbook,Blum,Schlosshauer}. Neglecting the system-bath correlations, tracing over the bath modes, and adopting the Markov approximation for bath correlation functions, we arrive at the Bloch-Redfield  (BR) master equation for the reduced density matrix of the V-system \cite{AgarwalMenon01,Scully06,prl14,Amr16}
\begin{align}\label{BRequations}
\dot\rho_{ii} &= -(\textit{r}_{i}+\gamma_{i}) \rho_{ii} + \textit{r}_{i} \rho_{cc} - p(\sqrt{\textit{r}_{a} \textit{r}_{b}}+\sqrt{\gamma_{a} \gamma_{b}}) \rho^R_{ab}\\
\dot\rho_{ab} &= -\frac{1}{2}(\textit{r}_{a}+\textit{r}_{b}+\gamma_{a}+\gamma_{b}) \rho_{ab} -i\rho_{ab} \Delta \nonumber \\&+ \frac{p}{2}\sqrt{\textit{r}_{a} \textit{r}_{b}} (2 \rho_{cc}-\rho_{aa}-\rho_{bb})
-\frac{p}{2}\sqrt{\gamma_{a} \gamma_{b}}(\rho_{aa}+\rho_{bb} )
\end{align}
where $a$, $b$, and $c$ are the system's energy eigenstates, the two-photon coherence $\rho_{ab} = \rho^{R}_{ab} +i \rho^{I}_{ab}$ is given as a sum of its real and imaginary parts, and we have used the conservation of probability condition to express $\rho_{aa}= 1-\rho_{bb}-\rho_{cc}$. 

The BR equations are parametrized by the excited-state energy splitting $\Delta = \omega_{ab}$ (see Fig. 1), the system-bath coupling parameters  $\gamma_i$ ($i = a,b$) which determine the rate of spontaneous decay into the vacuum  modes of the bath, and the (pseudo)thermal pumping rates $r_i=\gamma_i \bar{n}$ \cite{Keitel05}, where $\bar{n}$ is the effective occupation number of thermal modes at  the transition frequency $\omega_{ac}$ (see Fig. 1). In thermal equilibrium,  $\bar{n}=(1-e^{\beta\omega_0})^{-1}$, where $\beta=1/k_BT$, $T$ is the temperature of the bath, and $k_B$ is Boltzmann's constant. An important parameter $p= \frac{\vec \mu_{ac}.\vec \mu_{bc}}{|\vec \mu_{ac}| |\vec \mu_{bc}|}$ quantifies the alignment of the transition dipole moment vectors  $\vec{\mu}_{ac}$ and $\vec{\mu}_{bc}$ \cite{AgarwalMenon01,Scully06,prl14,Amr16}. We will show that the solutions of the BR equations tend to be extremely sensitive to the value of $p$. Note that for $p=0$, the BR equations reduce to the standard Pauli rate equations, which give coherence-free dynamics \cite{AgarwalMenon01,Scully06,prl14,Amr16}, We will therefore focus on the non-trivial case of $p \neq 0$.

The BR quantum master equations (\ref{BRequations}) generally describe the dynamics of the V-system interacting with stochastic bosonic fields, such as photons or phonons \cite{BPbook}. Here, we will consider the BR equations in a quantum optical context, relevant to the incoherent light excitation of quantum heat engines in the strong pumping limit,  $\bar{n}\gg 1$. We can then identify $\gamma_{i}=\frac{\omega_{ci}^{3} |\vec{\mu}_{ci}|^{2}} {3\pi \epsilon_{0} \hbar c^{3}}$ with the spontaneous emission rate  of the excited level $i = a, b$.
Further, $\textit{r}_{i}= B_{i} W(\omega_{ci})$ are the incoherent pumping rates of $\ket{c}$ $\leftrightarrow$ $\ket{i}$ transitions with $B_{i}=\frac{\pi |\vec{\mu}_{ci}|^{2}}{3 \epsilon_{0}\hbar^{2}}$ being the Einstein's $B$-coefficients and $W(\omega_{ci})$ is the intensity of the incident blackbody radiation at the corresponding transition frequencies. Finally, $r_{i}= \bar{n} \gamma_{i}$ are the incoherent absorption rates defined in terms of the effective photon occupation number $\bar{n}=B_{i} W(\omega_{ci})/\gamma_i$ \cite{Scully06,prl14,Keitel05}, which is proportional to the pumping intensity.

A comment is in order regarding the validity of  the BR quantum mater equations in the strong-pumping limit. The {\it weak-coupling} assumption underlying the BR equations holds as long as the system-bath coupling (as quantified by the incoherent pumping rates $r$) is much smaller than the energy gap $\omega_{ac}$ between the ground and excited energy eigenstates (see Fig.~1).   This condition is well satisfied for typical optical frequencies ($\omega_{ac}\sim 10^6$ GHz \cite{BPbook}) and incoherent pumping rates  ($r_i=1-10^3$~GHz) corresponding to the effective photon occupation numbers $\bar{n}=10-10^3$ typically used in few-level models of quantum heat engines  \cite{Chin13,Chin16,Scully11,Scully13}. Thus,  the strong-pumping condition $r\gg \gamma$  remains valid in the weak-coupling limit. In contrast, the Markovian assumption is expected to break down at very large pumping rates approaching the inverse bath correlation times $1/\tau_c$. In this limit, the BR equations remain valid as long as $r_i\ll 1/\tau_c$.  For incoherent pumping with solar light ($\tau_c\sim 1.3$ fs), this condition implies $\bar{n}\ll 1/(\tau_c\gamma_i)\simeq 10^6$, which is much larger than the effective photon occupation numbers considered here ($\bar{n}=10^{2}-10^3$). Following previous theoretical work \cite{Scully11,Scully13,Chin13,Chin16}, we neglect multiphoton transitions originating from the excited states of the V-system.


Here we consider the case of a symmetric V-system, where $\gamma_{a} = \gamma_{b} = \gamma$,   $\textit{r}_{a} = \textit{r}_{b} = \textit{r}$, and hence  $\rho_{aa}(t) = \rho_{bb}(t)$ \cite{prl14}.  The imposed symmetry simplifies the analytical solution of the BR equations to a great extent, while retaining the essential features of the dynamics \cite{prl14,Amr16}.
The BR master equations for the symmetric V-system (\ref{BRequations}) can be expressed in matrix-vector form 
\begin{equation}\label{BRmatrix}
\dot{\bm{x}}(t)=\mathbf{A}\bm{x}(t)+\bm{d},
\end{equation}
where ${\bm{x}}(t)=[\rho_{aa}(t),   \rho_{ab}^{R}(t),\rho_{ab}^{I}(t)]^T$ is the  state vector in the Liouville representation, where the elements of a $N\times N$ density matrix are represented by a vector of dimension $N^2$ \cite{Blum},  and $\bm{d}= [r,   p{r}, 0]^T$ is the driving vector.  Note that the state vector excludes the ground-state population and the one-photon coherences $\rho_{ac}$ and $\rho_{bc}$, which evolve independently  \cite{AgarwalMenon01}. The coefficient matrix $\mathbf{A}$ in Eq. (\ref{BRmatrix}) is given by
\begin{equation}\label{Amatrix}
\mathbf{A} = \begin{bmatrix}
    -(3\textit{r}+\gamma) & -p(\textit{r}+\gamma) & 0 \\
    -p(3\textit{r}+\gamma) & -(\textit{r}+\gamma) & \Delta  \\
       0 & -\Delta & -(\textit{r}+\gamma) \\
\end{bmatrix}.
\end{equation}

The general solution of the system of inhomogeneous differential equations (\ref{BRmatrix}) may be obtained as \cite{BoyceDiPrima}
\begin{equation}\label{Duhamel}
\bm{x}{(t)}=e^{\mathbf{A}t}\bm{x}_{0}+\int_{0}^{t}  ds e^{\mathbf{A}(t-s)} \bm{d} (s),
\end{equation}
where $\bm{x}_0$ specifies the initial conditions for the density matrix, and $\bm{d}(s)$ is the driving vector defined above. Since our interest here is in the generation of noise-induced Fano coherences by incoherent driving,  we choose a coherence-free initial state $\rho_{cc}(t=0)=1$, or  $\bm{x}_0 = (0, 0, 0)^T$, corresponding to the V-system initially in the ground state.
The exponent of matrix $\mathbf{A}$ in Eq. (\ref{Duhamel}) and the density matrix dynamics $\bm{x}{(t)}$ can be evaluated analytically in the limit $\bar{n}\gg1$ by expanding the matrix elements in the small parameter $x=1/\bar{n}$ as described in the Appendix.



\subsection{Dynamical regimes}

The behavior of the general solution of the BR equations (\ref{Duhamel}) is determined by the eigenvalue spectrum  $\lambda_k$ of the coefficient matrix $\mathbf{A}$. While the spectrum can be obtained analytically as described  below and in the Appendix, its general features can be understood by  examining the  discriminant $D$ of the characteristic equation for $\mathbf{A}$ 
\begin{equation}\label{D}
D= B^{3}+\left[C-\frac{3}{2}A(B+A^{2})\right]^{2},
\end{equation}
where
\begin{align}\label{ABC}\notag
A &=\frac{1}{3}(5r+3\gamma),\\ \notag
B &= \frac{1}{3}(r+\gamma)[\Delta^{2}+(r+\gamma)+(2-p^{2})(3r+\gamma)]-A^{2}, \\
C &= \frac{1}{2}(3r+\gamma)[\Delta^{2}+(1-p^{2})(r+\gamma)^{2}]+A^{3}
\end{align}
The above expressions are valid for all $p$.


Depending on the sign of $D$, three dynamical regimes can be distinguished:

\begin{enumerate}

\item 
{\it Underdamped regime}  ($D>0$).
If the discriminant (\ref{D}) is positive, one of the eigenvalues of $\mathbf{A}$  is real  and the other two eigenvalues are complex. The corresponding normal modes include an exponentially decaying eigenmode and two  oscillating eigenmodes. Using the analogy with the damped harmonic oscillator \cite{Amr16}, we will refer to this regime as {\it underdamped.}

\item 
{\it Overdamped regime} ($D<0$).
If the discriminant (\ref{D}) is negative, all of the eigenvalues $\lambda_k$ are real with $\text{Re}(\lambda_k) <0$, and thus all normal modes decay exponentially. Following our previous work \cite{Amr16}, we will refer to this regime as {\it overdamped}.

\item
{\it Critical regime} ($D=0$).
If the discriminant (\ref{D}) vanishes, all of the eigenvalues $\lambda_k$ are real with at least two of them being equal. This is the {\it critical regime} \cite{Amr16},  which  marks a transition between the underdamped and overdamped regimes.

\end{enumerate}

To classify the dynamical regimes of the strongly driven V-system, we therefore need to identify the regions of the parameter space where the discriminant (\ref{D}) takes on positive and negative values. As shown in the Appendix, the discriminant can be expressed as a polynomial function of the occupation number  $\bar{n}={r}/{\gamma}$
\begin{equation}\label{Dfinal}
D=\frac{\gamma^{6}}{108}  \sum\limits_{k=0}^{6} d_{k} \bar{n}^{k}
\end{equation}
where the coefficients $d_k$ depend  on the ratio $\Delta/\gamma$ of the excited-state splitting to the radiative decay rate and the transition dipole alignment factor $p$.

Figure \ref{fig:Discriminant1D} shows the lines of zero discriminant separating the overdamped ($D>0$) from underdamped ($D<0$) regimes as a function of the average photon occupation number $\bar{n}$ and the excited-state energy splitting $\Delta/\gamma$ for selected values of $p$. A contour plot of the discriminant is shown in  Fig. \ref{fig:Discriminant2D}.
We observe that when both  $\bar{n}$ and $\Delta/\gamma$ are large, the solution of the equation $D=0$ is given by the straight line $\Delta/\gamma = f(p)\bar{n}$.  It can be shown analytically (see the Appendix)  that the slope of the line $f(p)$ is a function of $p$ only. The slope function $f(p)$ plotted in Fig.~\ref{fig:f_of_p}(a) increases monotonically from zero to 0.6 as $p$ is varied between 0 and 1.


As illustrated in Fig.~\ref{fig:Discriminant2D}(a),  the discriminant is positive in the underdamped region above the zero-$D$ lines, where the coherences exhibit damped oscillations. Below the $D=0$ lines, the sign of $D$ changes from positive  to negative and the V-system enters the overdamped regime, with coherences evolving monotonously as a function of time.
Since, as shown above,  the zero-$D$ lines are described by $\Delta/\gamma = f(p)\bar{n}$ at large $\bar{n}$ and $\Delta/\gamma$, the dynamical regimes of the strongly driven V-system can be classified based on a single dimensionless parameter $\Delta/(\bar{n}\gamma)=\Delta/r$. The overdamped regime is defined by the condition $\Delta/(\bar{n}\gamma) < f(p)$ whereas the underdamped regime is defined by $\Delta/(\bar{n}\gamma)>f(p)$. For perfectly aligned transition dipole moments, we have $f(1)=0.6$ [see Fig.~\ref{fig:f_of_p}(a)] and the overdamped regime occurs for $\Delta/(\bar{n}\gamma)<0.6$. As the transition dipoles get out of alignment, the function $f(p)$ decreases, and smaller values of $\Delta$ are needed to reach the overdamped regime for a given $\bar{n}$. For instance, at $p=1/2$ the overdamped regime is reached for $\Delta/(\bar{n}\gamma)<0.16$ as illustrated  in Fig.~\ref{fig:Discriminant1D}(a), which shows that the slopes of the $D=0$ lines decrease proportionally to $p$.


Figures \ref{fig:Discriminant1D}(a) and \ref{fig:Discriminant1D}(b) show that the overdamped regime becomes progressively more widespread with increasing the pumping intensity $\bar{n}$. For large values of $p\simeq 1$ and $\bar{n}\gg 10$ of interest here, the underdamped regime is reached only at very large excited-state splittings ($\Delta/\gamma \gg 10$). In contrast, incoherent excitation of large molecules  with dense spectra of rovibrational levels \cite{prl14} and quantum heat engines \cite{Scully11,Scully13} typically occurs in the small level spacing regime $\Delta/\gamma \ll 1$. This is the regime we will consider in the remainder of this paper.  



As shown in Figs.~\ref{fig:Discriminant1D}(b), the zero-$D$ lines  approach constant values $\Delta/p\gamma$ in the weak pumping limit ($\bar{n}\to 0$). This implies that in this limit, the boundary between the overdamped and underdamped dynamical regimes is defined by the condition $\Delta/p\gamma = 1$, which is consistent with our previous results \cite{prl14,Amr16}.
It worth observing that the zero-$D$ lines in Fig.~\ref{fig:Discriminant1D}(b)  curve downward as $\bar{n}$ increases from zero to $\bar{n}\sim 0.01$. The reason for this is that the linear term ($d_{1} \bar{n}$) in Eq.~(\ref{Dfinal}) becomes negligible compared to the zeroth and second-order terms and the discriminant is given by $D = \frac{\gamma^{6}}{108} (d_{0}+d_{2} \bar{n}^{2})$. At higher values of $\bar{n}\sim 0.1$, the zero-$D$ lines reach a minimum and then start to approach their large-$\bar{n}$ limiting values as discussed above.


\subsection{Eigenvalues and coherence lifetimes}
As discussed in Sec. IIA, in order to obtain the general solution of the BR equations (\ref{Duhamel}), it is necessary to find the exponent of the coefficient matrix $\mathbf{A}$. To this end, we first diagonalize $\mathbf{A}$ to obtain the eigenvalues $\lambda_k$, which give the inverse lifetimes (or decay rates) of the corresponding normal modes $\bm{V}_k$ \cite{prl14,Amr16}.  Expanding the characteristic equation for $\mathbf{A}$ in terms of the small parameter $x = {\gamma}/{r}= {1}/{\bar{n}}$ (see the Appendix) we obtain the eigenvalues as  
\begin{equation}\label{lambda_exp}
\lambda_{j}=r\sum_{k=0}^{8}z_{jk}x^{k}, \quad (j=1,2,3).
\end{equation}
This expansion (\ref{lambda_exp}) is valid for  $x\le 0.01$ and $p>0.1$ (for $\Delta / \gamma < 1$) and $0.89 < p < 1$ (for $\Delta / \gamma > 1$).



In the overdamped regime, where $\frac{\Delta}{\gamma \bar{n}} \ll  f(p)$, the  expansion (\ref{lambda_exp}) converges rapidly. Keeping the lowest-order terms, we find to excellent accuracy
\begin{equation}\label{expansion3}
\lambda_{j}=r[z_{j0}+z_{j1} x + z_{j2} x^{2}],
\end{equation}
where the expressions for the coefficients $z_{jk}$ in terms of the system parameters $p$, $\Delta/\gamma$, and $\bar{n}$ are (see the Appendix)
\begin{align}\label{zj0and1}\notag
z_{j0}(p) &=\sqrt[3]{\frac{(8+27p^{2})}{27}+\sqrt{\frac{p^{4}(1+3p^{2})}{3}}i}\\ \notag
z_{j1}(p) &=-1-\frac{\alpha_{j}}{3K}4p^{2}+\left[\frac{\alpha_{j}}{3K}\left(\frac{4}{3}+3p^{2}\right)-\beta_{j}K\right]v_{1},\\
z_{j2}(p) &=f_{j1}(p)\left(\frac{\Delta}{\gamma}\right)^{2}+f_{j2}(p),
\end{align}
and the parameters $K$, $v_{1}$ and $f_{jk}(p)$ are listed in Tables 1, 2, and 4 of the Appendix. Here $\alpha_{j}$ and $\beta_{j}$ are the cube roots of unity with values $(\alpha_{1}, \beta_{1})= (1, 1)$, $(\alpha_{2}, \beta_{2})= (\omega^{2}, \omega)$, $(\alpha_{3}, \beta_{3})= (\omega, \omega^{2})$ with $\omega = \frac{(-1+i \sqrt{3})}{2}$ and $\omega^{2}= \frac{(-1 - i \sqrt{3})}{2}$. Note that since these parameters depend on $p$ only, the coefficients $z_{j1}$ and $z_{j0}$ are  independent of the ratio of the excited-state splitting to the radiative decay rate ${\Delta}/{\gamma}$. 
In contrast, the coefficient of $x^2$ in Eq. (\ref{expansion3}) carries an explicit quadratic dependence on  $\Delta/\gamma$.



In order to compare the relative importance of the different terms in Eq.~(\ref{expansion3}), we plot in Fig.~\ref{fig:zjk_terms} the $p$ dependence of  $\ln|z_{jk}|$  for $k =0$ - 2.  Figures~\ref{fig:zjk_terms}(a) and \ref{fig:zjk_terms}(c) show that $z_{10}$ provides the dominant contribution to $\lambda_1$ and $z_{30}$ provides the dominant contribution to $\lambda_3$  for all $p$, and we can thus approximate
\begin{equation}\label{zj0dominant}
\lambda_{j}=r z_{j0}=(\gamma z_{j0}) \bar{n}, \enspace  (j=1,3)
\end{equation}
where $z_{j0}(p)$ are given by Eq. (\ref{zj0and1}). The scaling behavior given by Eq. (\ref{zj0dominant}) is illustrated in Fig.~\ref{fig:eigenvalues_vsDelta}(b), which shows that the  eigenvalues $\lambda_1$ and $\lambda_3$ are independent of ${\Delta}/{\gamma}$ {\it regardless of the value of $p$}. 

 Remarkably, however, this is not the case for the eigenvalue $\lambda_2$: As shown  in Fig.~\ref{fig:zjk_terms}(b) there is a critical value of $p=p_c$ at which the curves $z_{20}(p)$ and $z_{22}(p) x^{2}$ cross and the relative contributions of the different terms to $\lambda_2$  change dramatically. At $p < p_{c}$, $z_{20}(p)$  is the dominating term so  $\lambda_{2}$ scales in the same way as the other eigenvalues (12). For $p > p_{c}$, the leading term is $z_{22} x^{2}$ and hence the scaling of $\lambda_{2}$ with $\Delta/\gamma$ is, quadratic the same as that of $z_{22}$ [see Eq.  (\ref{zj0and1})]. 
The critical value of $p$ depends on $\Delta/\gamma$ and $\bar{n}$ and ranges from 0.995 to 1.0 for $\bar{n}=10^{3}$ and $\Delta / \gamma = 10^{2} - 10^{-2}$ [See Fig. 4(b)]. 
The remarkable sensitivity of the second eigenvalue to the transition dipole alignment parameter $p$  shown in Fig.~\ref{fig:zjk_terms}(c)  leads to qualitatively different population and coherence dynamics for $p<p_c$ and $p>p_c$ as shown below.



As shown in Fig.~\ref{fig:zjk_terms}(b), the quadratic contribution to the second eigenvalue is much larger than the linear and constant terms, and hence $\lambda_{2} \simeq rz_{j2}({\gamma}/{r})^2$. Combining Eqs.~(\ref{expansion3}) and (\ref{zj0and1}) and noting that $f_{22}(p)\rightarrow 0$ for $p > p_{c}$, we find
\begin{equation}\label{lambda2expr}
\lambda_{2}= \frac{\gamma}{\bar{n}} f_{21}(p)\left(\frac{\Delta}{\gamma}\right)^{2}
\end{equation}
The distinct quadratic scaling of $\lambda_{2}$ with $\Delta/\gamma$  is illustrated in Fig.~\ref{fig:eigenvalues_vsDelta}(a). The function $f_{21}(p)$ (see the Appendix) increases monotonously approaching the value $-0.749$ in the limit $p\to 1$.
 As the second eigenvalue gives  the decay rate of the real part of the coherence (see Sec. IIA) the coherence lifetime $\tau_c = 1/|\lambda_2|$ is given by
\begin{equation}\label{tauCoherence}
\tau_{c} = 1.34\frac{\bar{n}}{\gamma} \left(\frac{\Delta}{\gamma}\right)^{-2} \qquad (p>p_c).
\end{equation}
The linear scaling of the coherence lifetime with the  incoherent pumping intensity $\bar{n}$ is a direct consequence of the second eigenvalue being dominated by a single $z_{22}x^2$ contribution as shown in Fig.~\ref{fig:zjk_terms}(b). This characteristic scaling occurs only for $p>p_c$ with $p_c = 1-\epsilon$ very close to unity [typical values of  $\epsilon$ range from  $5\times 10^{-11}$ to $5\times 10^{-3}$  for $\bar{n} = 10^{3}$ as shown in Fig.~4(b). Thus, the V-system with nearly parallel transition dipole moments can  exhibit very long coherence times in the strong pumping limit.  

For subcritical transition dipole alignment ($p<p_c$), the coherence lifetime becomes
\begin{equation}\label{tauCoherenceLess}
\tau_{c} = \frac{1}{\gamma z_{20}(p)}  \bar{n}^{-1}  \qquad  (p<p_c).
\end{equation}
 The coherence lifetimes thus become shorter with increasing the pumping intensity, in stark contrast with the situation in the  supercritical regime ($p>p_c$), where the lifetimes increase linearly  with $\bar{n}$ (\ref{tauCoherence}).


It is instructive to compare the coherence time of the V-system in the strong-pumping  and weak-pumping regimes. In the small level spacing regime ($\Delta / \gamma < 1$), the coherence time under weak pumping $\tau_{c}^{WP} = \frac{2}{\gamma} \left(\frac{\Delta}{\gamma}\right)^{-2}$  \cite{prl14,Amr16}  exhibits the same $(\Delta/\gamma)^{-2}$ scaling as in the strong-pumping regime \cite{prl14,Amr16}, becoming longer as the excited-state energy gap $\Delta$ narrows down. The ratio of the coherence times in the strong and weak-pumping limits is thus, for $p>p_c$
\begin{equation}\label{tauCoherenceWeakPump}
\frac{\tau_{c}}{\tau_c^{WP}} \simeq 0.67 \bar{n}
\end{equation}
The enhancement of the coherence lifetime  under strong pumping ($n>1.5$) may facilitate the experimental observation of the noise-induced coherences in atomic systems \cite{AmrCa17}.  


  We note in passing that nearly perfect alignment of the transition dipole moments ($p>p_c$) is an essential condition for the longevity of noise-induced coherences in the strong pumping regime.  In practice, since $p_c$ is very close to unity, this condition is equivalent to the requirement  $p=1$.
The coherence  lifetime exhibits two dramatically different scalings with the pumping intensity $\bar{n}$. For $p<p_c$, the lifetime decreases with $\bar{n}$, whereas for $p>p_c$ the opposite trend is observed.  This means that in the strong-pumping limit, even slightest misalignment of the transition dipole moments will destroy long-lived coherent dynamics. 





\section{Population and coherence dynamics}

\subsection{Analytic solutions in the overdamped regime [$\Delta/(\bar{n}\gamma) < f(p)$] }

Having classified the dynamical regimes of the strongly driven V-system and analyzed the relevant eigenmodes, we now turn to the time evolution of the density matrix elements. From Eq.~(\ref{Duhamel}), we obtain [24]
\begin{equation}\label{DMdynamics}
\rho_{ij}(t) =\frac{r}{\det(\mathbf{M})} \sum_{k=1}^{3}\frac{(e^{\lambda_{k}t}-1)}{\lambda_{k}}V_{k,n(i,j)}(T_{k1}+pT_{k2}).
\end{equation} 
where $n(i,j)$ = 1 for the excited-state population ($i=j=a$) and $n(i,j)=2(3)$ for the real (imaginary) part of the two-photon coherence ($i=a,j=b$). The density matrix elements in Eq.~(\ref{DMdynamics}) are expressed as a linear combination of exponentially decaying terms, weighted with the  components $V_{kj}$ of eigenvectors of $\mathbf{A}$ (which form the fundamental matrix $\mathbf{M}$) and the elements of its adjoint matrix $T_{kj}$, which depend on $p$ only. 

To express the density matrix dynamics in terms of the physical parameters $\Delta/\gamma$, $\bar{n}$, and $p$, we evaluate the eigenvector components $V_{kj}$ and $T_{kj}$ in Eq.~(\ref{DMdynamics})  as described in the Appendix and use the resulting expressions in Eq.~(\ref{Duhamel}) to obtain for $p>p_{c}$
\begin{equation}\label{GeneralSol1}
\begin{split}
\rho_{aa}(t) &= \frac{1}{[T_{1}(p)+T_{2}(p)\frac{1}{\bar{n}^{2}}(\frac{\Delta}{\gamma})^{2}]}\biggl{\{}\left[A_{1}+A_{2}\frac{1}{\bar{n}^{2}}\left(\frac{\Delta}{\gamma}\right)^{2}\right] \frac{1-e^{-\gamma|z_{10}(p)|\bar{n}t}}{ |z_{10}(p)|}
\\ &+\left[ A_{3}+A_{4}\frac{1}{\bar{n}^{2}}\left(\frac{\Delta}{\gamma}\right)^{2}\right]\bar{n}^{2}\left(\frac{\Delta}{\gamma}\right)^{-2} \frac{1-e^{-\gamma|f_{21}(p)|\frac{1}{\bar{n}}(\frac{\Delta}{\gamma})^{2}t}}{|f_{21}(p)|} +
A_{5}\frac{1}{\bar{n}^{2}}\left(\frac{\Delta}{\gamma}\right)^{2}\frac{1-e^{-\gamma|z_{30}(p)|\bar{n}t}}{ |z_{30}(p)|}\biggr{\}}
\end{split}
\end{equation}
\begin{equation}\label{GeneralSol2}
\begin{split}
\rho^{R}_{ab}(t)&= \frac{1}{[T_{1}(p)+T_{2}(p)\frac{1}{\bar{n}^{2}}(\frac{\Delta}{\gamma})^{2}]}\biggl{\{} \left[B_{1}+B_{2}\frac{1}{\bar{n}^{2}}\left(\frac{\Delta}{\gamma}\right)^{2}\right]\frac{1-e^{-\gamma|z_{10}(p)|\bar{n}t}}{ |z_{10}(p)|}
\\ &+\left[B_{3}+B_{4}\frac{1}{\bar{n}^{2}}\left(\frac{\Delta}{\gamma}\right)^{2}\right]\bar{n}^{2}\left(\frac{\Delta}{\gamma}\right)^{-2}\frac{1-e^{-\gamma|f_{21}(p)|\frac{1}{\bar{n}}(\frac{\Delta}{\gamma})^{2}t}}{|f_{21}(p)|}
+B_{5}\frac{1}{\bar{n}^{2}}\left(\frac{\Delta}{\gamma}\right)^{2}\frac{1-e^{-\gamma|z_{30}(p)|\bar{n}t}}{ |z_{30}(p)|} \biggr{\}}
\end{split}
\end{equation}
\begin{equation}\label{GeneralSol3}
\begin{split}
\rho^{I}_{ab}(t)&= \frac{1}{[T_{1}(p)+T_{2}(p)\frac{1}{\bar{n}^{2}}(\frac{\Delta}{\gamma})^{2}]} \frac{1}{\bar{n}}\left(\frac{\Delta}{\gamma}\right) \biggl{\{} \left[C_{1}+C_{2}\frac{1}{\bar{n}^{2}}\left(\frac{\Delta}{\gamma}\right)^{2}\right]\frac{1-e^{-\gamma|z_{10}(p)|\bar{n}t}}{ |z_{10}|} \\&+ \left[C_{3}+C_{4}\frac{1}{\bar{n}^{2}}\left(\frac{\Delta}{\gamma}\right)^{2}\right]\bar{n}^{2}\left(\frac{\Delta}{\gamma}\right)^{-2}\frac{1-e^{-\gamma|f_{21}(p)|\frac{1}{\bar{n}}(\frac{\Delta}{\gamma})^{2}t}}{|f_{21}(p)|}+
\left[C_{5}+C_{6}\frac{1}{\bar{n}^{2}}\left(\frac{\Delta}{\gamma}\right)^{2}\right]\frac{1-e^{-\gamma|z_{30}|\bar{n}t}}{ |z_{30}(p)|}\biggr{\}},
\end{split}
\end{equation}
where the eigenvalue components $z_{j0}(p)$ are given by Eqs.~(\ref{zj0and1}) and  the coefficients $A_i$, $B_i$, and $C_i$ plotted in Figs.~\ref{fig:ABCodd_terms} and \ref{fig:ABCeven_terms}  depend on $p$ only (see the Appendix for analytic expressions, which are rather cumbersome). The second term on the right-hand-side of Eq. (\ref{GeneralSol2}) represents the slowly decaying coherent mode discussed in Sec. IIB, which also manifests itself in the time evolution of excited-state populations (\ref{GeneralSol1}).  The lifetime of the coherent mode scales as $\bar{n}(\Delta/\gamma)^{-2}$, leading to arbitrarily long coherence lifetimes for small excited-state splittings. In contrast,  the first and third terms on the right-hand side of Eqs.~(\ref{GeneralSol1})-(\ref{GeneralSol3}) decay much faster, with lifetimes proportional to $1/\bar{n}$.  

To further simplify our analytic solutions (\ref{GeneralSol1})-(\ref{GeneralSol3}), we note that the coefficients $A_i$, $B_i$, and $C_i$ plotted in Figs.~\ref{fig:ABCodd_terms} and \ref{fig:ABCeven_terms} do not vary strongly with $p$ in the vicinity of $p=1$.
We can thus replace the coefficients  by their limiting values at $p\to 1$ to yield
\begin{equation}\label{GeneralSol1p1}
\begin{split}
\rho_{aa}(t) &= \frac{1}{[-4+1.33\frac{1}{\bar{n}^{2}}(\frac{\Delta}{\gamma})^{2}]}\biggl{\{}\left[-4 + 2.92 \frac{1}{\bar{n}^{2}}\left(\frac{\Delta}{\gamma}\right)^{2}\right] \frac{(1-e^{-4\gamma  \bar{n}t})}{4}
\\ &-\frac{(1-e^{-0.75 \frac{\gamma}{\bar{n}}(\frac{\Delta}{\gamma})^{2}t})}{3} +
0.44 \frac{1}{\bar{n}^{2}}\left(\frac{\Delta}{\gamma}\right)^{2} (1-e^{-\gamma \bar{n}t})\biggr{\}}
\end{split}
\end{equation}
\begin{equation}\label{GeneralSol2p1}
\begin{split}
\rho^{R}_{ab}(t)&= \frac{1}{[-4+1.33\frac{1}{\bar{n}^{2}}(\frac{\Delta}{\gamma})^{2}]}\biggl{\{} \left[-4+3.249\frac{1}{\bar{n}^{2}}\left(\frac{\Delta}{\gamma}\right)^{2}\right]\frac{(1-e^{-4 \gamma \bar{n}t})}{4}
\\ &+ (1-e^{-0.75 \frac{\gamma}{\bar{n}}(\frac{\Delta}{\gamma})^{2}t})
-0.89\frac{1}{\bar{n}^{2}}\left(\frac{\Delta}{\gamma}\right)^{2} (1-e^{-\gamma \bar{n}t}) \biggr{\}}
\end{split}
\end{equation}
\begin{equation}\label{GeneralSol3p1}
\begin{split}
\rho^{I}_{ab}(t)&= \frac{1}{[-4+1.33\frac{1}{\bar{n}^{2}}(\frac{\Delta}{\gamma})^{2}]} \frac{1}{\bar{n}}\left(\frac{\Delta}{\gamma}\right) \biggl{\{} \left[-1.33+0.99\frac{1}{\bar{n}^{2}}\left(\frac{\Delta}{\gamma}\right)^{2}\right]\frac{(1-e^{-4 \gamma \bar{n}t})}{4} \\& - (1-e^{-0.75 \frac{\gamma}{\bar{n}}(\frac{\Delta}{\gamma})^{2}t})+
\left[1.33-0.25\frac{1}{\bar{n}^{2}}\left(\frac{\Delta}{\gamma}\right)^{2}\right] (1-e^{-\gamma \bar{n}t}) \biggr{\}},
\end{split}
\end{equation}



For subcritical transition dipole alignment ($p<p_{c}$) the population and coherence dynamics take the form
\begin{equation}\label{GeneralSol1p}
\begin{split}
\rho_{aa}(t) &= \frac{1}{[T_{1}(p)+T_{2}(p)\frac{1}{\bar{n}^{2}}(\frac{\Delta}{\gamma})^{2}]} \biggl{\{} 
\left[ A_{1}+A_{2}\frac{1}{\bar{n}^{2}}\left(\frac{\Delta}{\gamma}\right)^{2}\right] \frac{1-e^{-\gamma|z_{10}|\bar{n}t}}{ |z_{10}|}
\\ &+\left[A_{3}+A_{4}\frac{1}{\bar{n}^{2}}\left(\frac{\Delta}{\gamma}\right)^{2}\right] \frac{1-e^{-\gamma|z_{20}|\bar{n}t}}{ |z_{20}|}+
A_{5}\frac{1}{\bar{n}^{2}}\left(\frac{\Delta}{\gamma}\right)^{2}\frac{1-e^{-\gamma|z_{30}|\bar{n}t}}{ |z_{30}|} \biggr{\}}
\end{split}
\end{equation}
\begin{equation}\label{GeneralSol2p}
\begin{split}
\rho^{R}_{ab}(t)&= \frac{1}{[T_{1}(p)+T_{2}(p)\frac{1}{\bar{n}^{2}}(\frac{\Delta}{\gamma})^{2}]}
\biggl{\{} \left[B_{1}+B_{2}\frac{1}{\bar{n}^{2}}\left(\frac{\Delta}{\gamma}\right)^{2}\right] \frac{1-e^{-\gamma|z_{10}|\bar{n}t}}{ |z_{10}|}
\\ &+\left[B_{3}+B_{4}\frac{1}{\bar{n}^{2}}\left(\frac{\Delta}{\gamma}\right)^{2}\right]  \frac{1-e^{-\gamma|z_{20}|\bar{n}t}}{ |z_{20}|} + B_{5}\frac{1}{\bar{n}^{2}}\left(\frac{\Delta}{\gamma}\right)^{2} \frac{1-e^{-\gamma|z_{30}|\bar{n}t}}{ |z_{30}|} \biggr{\}}
\end{split}
\end{equation}
\begin{equation}\label{GeneralSol3p}
\begin{split}
\rho^{I}_{ab}(t)&= \frac{1}{[T_{1}(p)+T_{2}(p)\frac{1}{\bar{n}^{2}}(\frac{\Delta}{\gamma})^{2}]}\frac{1}{\bar{n}}\left(\frac{\Delta}{\gamma}\right) \biggr{\{} \left[(C_{1}+C_{2}\frac{1}{\bar{n}^{2}}\left(\frac{\Delta}{\gamma}\right)^{2}\right] \frac{1-e^{-\gamma|z_{10}|\bar{n}t}}{ |z_{10}|} \\ &+ \left[C_{3}+C_{4}\frac{1}{\bar{n}^{2}}\left(\frac{\Delta}{\gamma}\right)^{2}\right] \frac{1-e^{-\gamma|z_{20}|\bar{n}t}}{ |z_{20}|}) +
\left[C_{5}+C_{6}\frac{1}{\bar{n}^{2}}\left(\frac{\Delta}{\gamma}\right)^{2}\right] \frac{1-e^{-\gamma|z_{30}|\bar{n}t}}{ |z_{30}|} \biggr{\}}
\end{split}
\end{equation}
where, in contrast to the $p>p_c$ case considered above, all of the exponential terms on the right-hand side  scale linearly with $\bar{n}$ and are  independent of $\Delta/\gamma$ (see Sec. IIB). The coherence dynamics given by Eqs. (\ref{GeneralSol1p})-(\ref{GeneralSol3p}) is thus much more short lived than that observed for the case of nearly parallel transition dipole moments.

Figure~\ref{fig:dyn_largeDelta} compares our analytical results (\ref{GeneralSol1})-(\ref{GeneralSol3p}) with the time evolution of the populations and coherences obtained by numerical integration of the BR equations (\ref{BRequations}) for $\Delta/\gamma=10$. A sudden turn-on of incoherent pumping at $t=0$ initiates population transfer from the ground state to the excited eigenstates and generates two-photon coherences among them.  We observe excellent agreement between the analytical and numerical dynamics. 
 The excited-state populations grow monotonously to their steady-state values $\rho_{aa}/\rho_{gg}=1$ given by the Boltzmann distribution as expected for the solutions of the BR equations ~\cite{AgarwalMenon01}.  
  
   It follows from the analytical expressions for $\rho_{aa}$ and $\rho_{ab}$  [Eqs. (21)-(22)] that the time it takes for the populations to reach the steady state is the same as the lifetime of the long-lived coherent mode given by Eq. (\ref{tauCoherence}). Figures~\ref{fig:dyn_largeDelta}(a) and \ref{fig:dyn_largeDelta}(d) show that the timescale on which the V-system thermalizes with the bath depends strongly on the value of the transition dipole alignment factor $p$. For $p>p_c$, the coherence lifetime can be extremely long and full thermalization does not occur until after $t> \tau_c$ as shown in Fig.~\ref{fig:dyn_largeDelta}(a). 

   Importantly, at shorter times ($t<\tau_c$), instead of the expected coherence-free, canonical steady state,  we observe the formation of a long-lived {\it quasisteady  state} featuring significant coherences in the energy eigenstate basis.  The excited-state populations in the quasisteady state are suppressed compared to those in thermal equilibrium, as previously found for the weakly driven V-system \cite{prl14,Amr16}. The emergence of the long-lived, coherent quasisteady state is another remarkable aspect of the fully aligned V-system undergoing thermalization dynamics  with the bath. 
Finally, we note that no quasisteady states are formed for $p<p_c$, leading to more conventional thermalization dynamics shown in  Fig.~\ref{fig:dyn_largeDelta}(d), in contrast to the weakly driven V-system case, where the long-lived coherent states are present even for $p<p_c$ \cite{prl14,Amr16}.  The sharp transition between the different dynamical regimes as a function of $p$ is thus a unique feature of the strongly driven V-system.


The real and imaginary parts of the two-photon coherence shown in Figs.~\ref{fig:dyn_largeDelta}(b) and \ref{fig:dyn_largeDelta}(c) grow monotonously,  reaching a plateau on the timescale $t\sim 1/r=\bar{n}{\gamma}$  and then surviving for the duration $\tau_c$ of the coherence lifetime given by Eq.~(\ref{tauCoherence}).    By comparing Fig.~\ref{fig:dyn_largeDelta}(b) and Fig.~\ref{fig:dyn_largeDelta}(e), we observe that the coherences become much more short-lived for subcritical transition dipole alignment ($p<p_c$) in accordance with  Eqs. (\ref{tauCoherence}) and (\ref{tauCoherenceLess}).  In the limit $t \gg \tau_c$ the coherences decay to zero and the V-system reaches the expected thermal equilibrium state \cite{AgarwalMenon01}. 


\subsection{Analytic solutions for closely spaced levels [$\Delta/\gamma \ll 1$] }

As discussed in the previous section the analytical expressions (\ref{GeneralSol1}) - (\ref{GeneralSol3}) are valid in the overdamped regime defined by the condition $\Delta/(\bar{n} \gamma) <  f(p)$. Since $\bar{n}\gg 1$, the  condition $\Delta/(\bar{n} \gamma) <  f(p)$ does not necessarily imply that the  excited-state splittings should be small compared to the radiative decay rate ({\it i.e.} $\Delta/\gamma \ll 1$). As a result,  the strongly driven V-system can exhibit overdamped coherent behavior even when the excited-state level splitting is large compared to the natural linewidth ($\Delta/\gamma \gg 1$)  provided that $\Delta/\gamma < f(p)\bar{n}$.   Nevertheless, major simplifications are possible in the limit of closely spaced excited-state levels ($\Delta/\gamma \ll 1$),  which is of special interest for incoherent excitation of large molecules \cite{prl14}. It is also in this limit that the weakly driven V-system exhibits long-lived coherences  \cite{prl14,Amr16}.   

Neglecting the terms proportional to $(\Delta/\gamma)^2$ in the expressions for the population and coherence dynamics (\ref{GeneralSol1})-(\ref{GeneralSol3}) we obtain for nearly parallel transition dipole moments ($p>p_{c}$) 
\begin{align}\label{GenSolCloselySpaced}
\rho_{aa}(t) &= \frac{1}{T_{1}(p)}\biggl{\{} A_{1} \frac{1-e^{-\gamma|z_{10}(p)|\bar{n}t}}{ |z_{10}(p)|}
+A_{4}  \frac{1-e^{-\gamma|f_{21}(p)|\frac{1}{\bar{n}}\left(\frac{\Delta}{\gamma}\right)^{2}t}}{|f_{21}(p)|} \biggr{\}}\\  
%
\rho^{R}_{ab}(t) &= \frac{1}{T_{1}(p)}\biggl{\{} B_{1} \frac{1-e^{-\gamma|z_{10}(p)|\bar{n}t}}{ |z_{10}(p)|}
+B_{4} \frac{1-e^{-\gamma|f_{21}|\frac{1}{\bar{n}}\left(\frac{\Delta}{\gamma}\right)^{2}t}}{|f_{21}(p)|} \biggr{\}}\\ 
%
\rho^{I}_{ab}(t) &= \frac{1}{T_{1}(p)}\left(\frac{\Delta}{\bar{n}\gamma}\right) \biggl{\{} C_{1}\frac{1-e^{-\gamma|z_{10}(p)|\bar{n}t}}{ |z_{10}(p)|}
+C_{4} \frac{1-e^{-\gamma|f_{21}(p)|\frac{1}{\bar{n}}(\frac{\Delta}{\gamma})^{2}t}}{|f_{21}(p)|} +
C_{5} \frac{1-e^{-\gamma|z_{30}(p)|\bar{n}t}}{ |z_{30}(p)|}\biggr{\}}
\end{align}
where the quantities $z_{jk}$ and $f_{21}$ depend on $p$ only (see the Appendix and Sec. IIC above)  and $|f_{21}(p)| = 0.749$ in the limit $p \to 1$.

To further simplify our analytic solutions (27)-(29), we note that the coefficients $A_i$, $B_i$, and $C_i$ plotted in Figs.~\ref{fig:ABCodd_terms} and \ref{fig:ABCeven_terms} do not vary strongly with $p$ in the vicinity of $p=1$. Replacing the coefficients by their values at $p = 1$, we find
\begin{align}\label{GenSolCloselySpaced}
\rho_{aa}(t) &= \frac{1}{3} - \frac{1}{12}  \left( 3 e^{-4 \gamma \bar{n}t} + e^{-0.75 \frac{\gamma}{\bar{n}} (\frac{\Delta}{\gamma})^{2} t} \right) \\ 
%
\rho^{R}_{ab}(t) &= \frac{1}{4}  \left(e^{-0.75 \frac{\gamma}{\bar{n}} (\frac{\Delta}{\gamma})^{2} t} -e^{-4 \gamma \bar{n}t} \right) \\ 
\rho^{I}_{ab}(t) &= -\frac{1}{12}\left(\frac{\Delta}{\bar{n}\gamma}\right) \left( e^{- 4 \gamma \bar{n} t} + 3 e^{-0.75 \frac{\gamma}{\bar{n}} (\frac{\Delta}{\gamma})^{2} t} -
4 e^{-\gamma \bar{n} t} \right)
\end{align}
These expressions clearly establish the existence of  two vastly different timescales of coherent dynamics. At very short times \big($t < \frac{1}{4 \gamma \bar{n}}$\big) the real part of the coherence increases to its quasisteady value of $\frac{1}{4}$ as shown in Fig. 10(a). The coherence survives for a long time $\tau_{c}$ given by Eq. (14) before eventually decaying to zero. The imaginary part of the coherence is suppressed by the factor \big($\frac{\Delta}{\bar{n}\gamma} \ll 1$\big), as previously found in the weak-pumping limit [13,14].


For imperfectly aligned transition dipole moments ($p<p_{c}$) we find 
\begin{align}
\rho_{aa}(t) &= \frac{1}{T_{1}(p)} \biggl{\{} A_{1} \left(\frac{1-e^{-\gamma|z_{10}|\bar{n}t}}{ |z_{10}|}\right)+A_{3} \left(\frac{1-e^{-\gamma|z_{20}|\bar{n}t}}{ |z_{20}|}\right) \biggr{\}} \\
\rho^{R}_{ab}(t) &= \frac{1}{T_{1}(p)} \biggl{\{} B_{1}\left(\frac{1-e^{-\gamma|z_{10}|\bar{n}t}}{ |z_{10}|}\right)
+B_{3}\left(\frac{1-e^{-\gamma|z_{20}|\bar{n}t}}{ |z_{20}|}\right) \biggr{\}} \\
\rho^{I}_{ab}(t) &= \frac{1}{T_{1}(p)} \left(\frac{\Delta}{\bar{n} \gamma}\right) \biggl{\{} C_{1}\left(\frac{1-e^{-\gamma|z_{10}|\bar{n}t}}{ |z_{10}|}\right)+C_{3}\left(\frac{1-e^{-\gamma|z_{20}|\bar{n}t}}{ |z_{20}|}\right)+C_{5}\left(\frac{1-e^{-\gamma|z_{30}|\bar{n}t}}{ |z_{30}|}\right) \biggr{\}}
\end{align}
As shown in Figs. 10(d)-(f), these analytical solutions are in excellent agreement with numerical results. The population and coherence dynamics in the limit of small excited-state splitting is qualitatively similar to that  observed for large excited-state splittings discussed above. 



\section{Summary and conclusions}

We have studied the quantum dynamics of a three-level V-system interacting with a thermal environment in a previously unexplored regime, where incoherent pumping occurs much faster than spontaneous emission. This regime is characterized by a large number of thermal bath phonons ($\bar{n}\gg1$) at the excitation frequency,  and it is relevant for artificial solar light harvesting and the design of efficient quantum heat engines \cite{Chin13,Chin16,Scully11,Scully13}, of which the V-system is a key building block. 

As a primary tool to study the dynamics of the strongly driven V-system, we use non-secular BR equations, which provide a unified description of time-evolving populations and coherences in multilevel quantum systems interacting with a thermal bath. The non-secular description retains the population-to-coherence coupling terms proportional to the transition dipole alignment factor $p$, which are essential for a proper description of  noise-induced coherences \cite{prl14}.  By examining the discriminant of the characteristic  polynomial, we  classify the dynamical regimes of the strongly driven V-system into underdamped, overdamped, and critical (Sec.~IIA). For large excited-state splittings such that $\Delta/(\bar{n}\gamma) >f(p)$, where $f(p)$ is a universal function of $p$ plotted in Fig.~\ref{fig:f_of_p}, the two-photon  coherences show underdamped oscillations. In the overdamped regime of small level spacing [$\Delta/(\bar{n}\gamma) < 1$], the coherences evolve monotonously as a function of time. 
A remarkable dynamical effect which occurs in this regime is the formation of long-lived, coherent quasisteady states with  lifetimes $\tau_c=1.34(\bar{n}/\gamma)({\Delta}/\gamma)^{-2}$ that can be arbitrarily long in V-systems with vanishingly small level splittings. As illustrated in  Fig.~\ref{fig:dyn_largeDelta}, the quasi-steady states strongly affect the time evolution of the density matrix elements,  enhancing the lifetime of  two-photon coherences and slowing down the approach of excited-state populations to thermodynamic equilibrium.  The quasisteady states only form  when the transition dipole moments of the V-system are nearly perfectly aligned,
  in contrast to the behavior of the weakly driven V-system, which can display long-lived coherence dynamics for $p<p_c$ \cite{prl14,Amr16}.

We further show that in the overdamped regime, the solutions of the BR equations can be represented analytically as a sum of three exponentially decaying terms (\ref{GeneralSol1})-(\ref{GeneralSol3}).  The behavior of the solutions depends strongly on the transition dipole alignment factor $p$. For   $p>p_c$, the long-lived coherent mode emerges, whereas for $p<p_c$ all modes have comparable lifetimes, which scale as $1/\bar{n}$. Particularly simple expressions (\ref{GenSolCloselySpaced}) are obtained in the limit of small level spacing $\Delta/\gamma \ll 1$. All of the expressions are in excellent agreement with numerical solutions of the BR equations (Figs. 9 and 10).


Finally, we consider the question of how the long-lived noise-induced coherent effects predicted here could be observed in the laboratory. Such an observation would require an atomic or molecular V-system with nearly parallel transition dipole moments ($p>p_c$) driven by a bright $(\bar{n}\gg1)$ source of incoherent radiation. For the latter, one can use concentrated solar light (for which $\bar{n}\simeq 10^4$  can be achieved at  typical optical frequencies \cite{Chin16,Scully11,Scully13}) or broadband laser radiation \cite{AgarwalMenon01}.  

 The  requirement of nearly perfectly aligned transition dipoles ($p>p_c$) is more restrictive, since the overwhelming majority of electric dipole transitions to nearly degenerate (V-system-like) excited atomic states tend to have $p<0$ \cite{AgarwalMenon01,Scully92,prl14}.
  To bypass this requirement,  we  consider incoherent excitation by a linearly polarized blackbody radiation,  for which  transitions to nearly degenerate upper levels with different $m_J$ exhibit Fano interference  \cite{AgarwalMenon01,AmrCa17}.
 
 In future work, we intend to explore the possibility for experimental observation of Fano coherences with highly excited Rydberg atoms. As a consequence of  their exaggerated  transition dipole moments, Rydberg atoms couple strongly to blackbody radiation, and can therefore be used as an attractive experimental platform to study  noise-induced coherence effects  \cite{RydbergBook,RydbergCMB}. Consider, {\it e.g.}, the $65s$ Rydberg state of atomic Rb interacting with thermal blackbody radiation at $T=300$~K \cite{RydbergCMB}. The energy splitting $\omega_{ac}$ between the initial $65s$ state and the nearby $m_J$ components of the $65p$ state (forming the Rydberg V-system) is 0.44 cm$^{-1}$. The average number of thermal photons at this transition frequency is $\bar{n}\sim 400$ \cite{RydbergBook}, putting the Rydberg V-system in the strong pumping regime. The splitting between the different $m_J$ components of the $65p$ state  can be tuned by an external magnetic field to vary the ratio $\Delta/\bar{n}\gamma$, providing access to the different regimes of noise-induced coherent dynamics studied in this work.
 



\newpage


\clearpage
\newpage

\begin{figure}[t!]
\captionsetup{singlelinecheck = false, format= hang, justification=raggedright, font=footnotesize, labelsep=space}
\centering
\includegraphics[height=6.0cm, width=12.0cm]{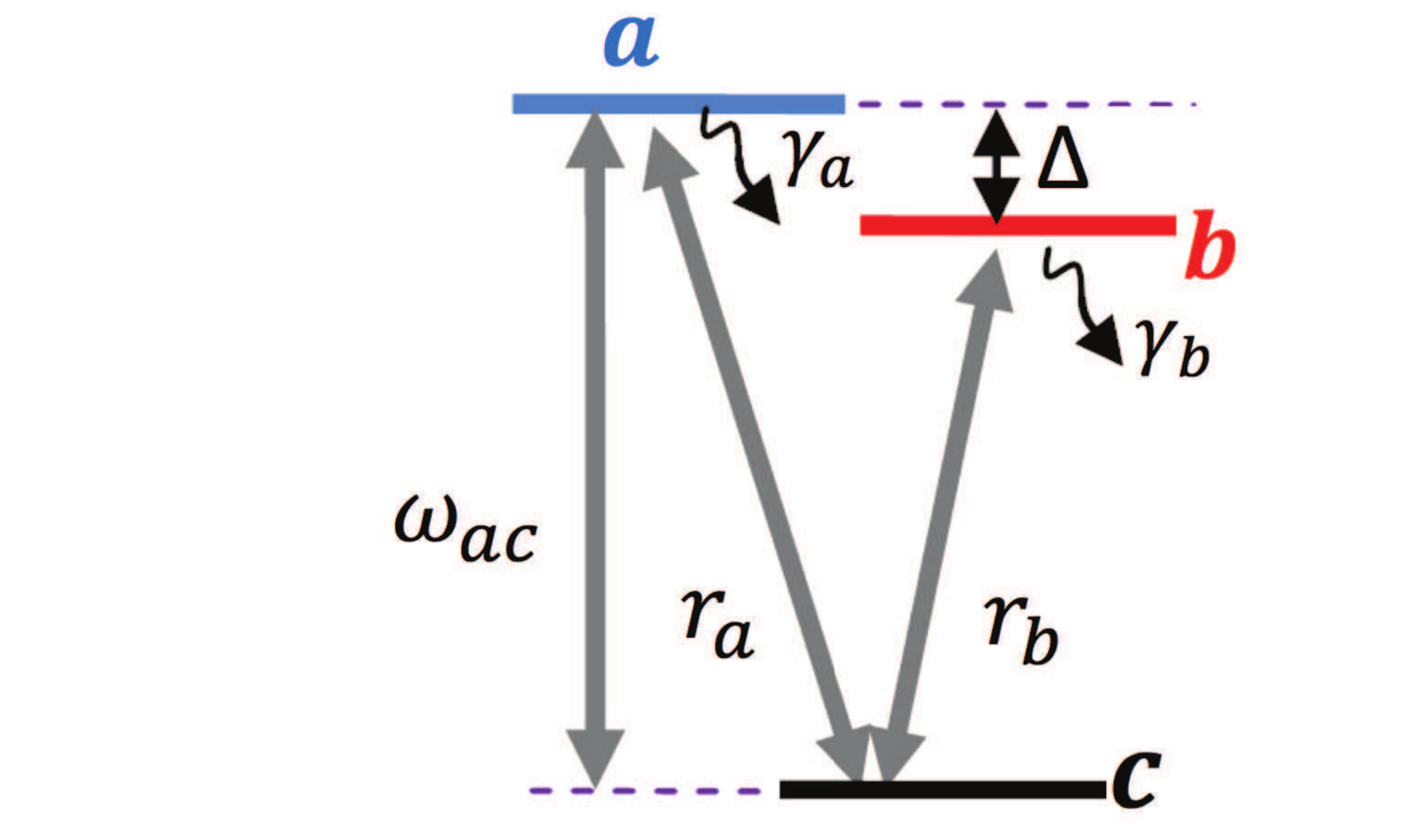}
\caption{A schematic depiction of the three-level V-system characterized by the excited-state energy splitting  $\Delta = \omega_{ab}$, the rates of spontaneous decay into the vacuum modes of the bath $\gamma_i$ ($i = a,b$), and the thermal pumping rates $r_i=\gamma_i \bar{n}$. The factor $p= {\vec \mu_{ac}\cdot \vec \mu_{bc}}/{|\vec \mu_{ac}| |\vec \mu_{bc}|}$ quantifies the alignment of the  transition dipole moment   vectors  $\vec{\mu}_{ac}$ and $\vec{\mu}_{bc}$.}
\label{fig:Vsystem}
\end{figure}

\begin{figure}[t!]
\captionsetup{singlelinecheck = false, format= hang, justification=raggedright, font=footnotesize, labelsep=space}
     \includegraphics[width=1.05\textwidth]{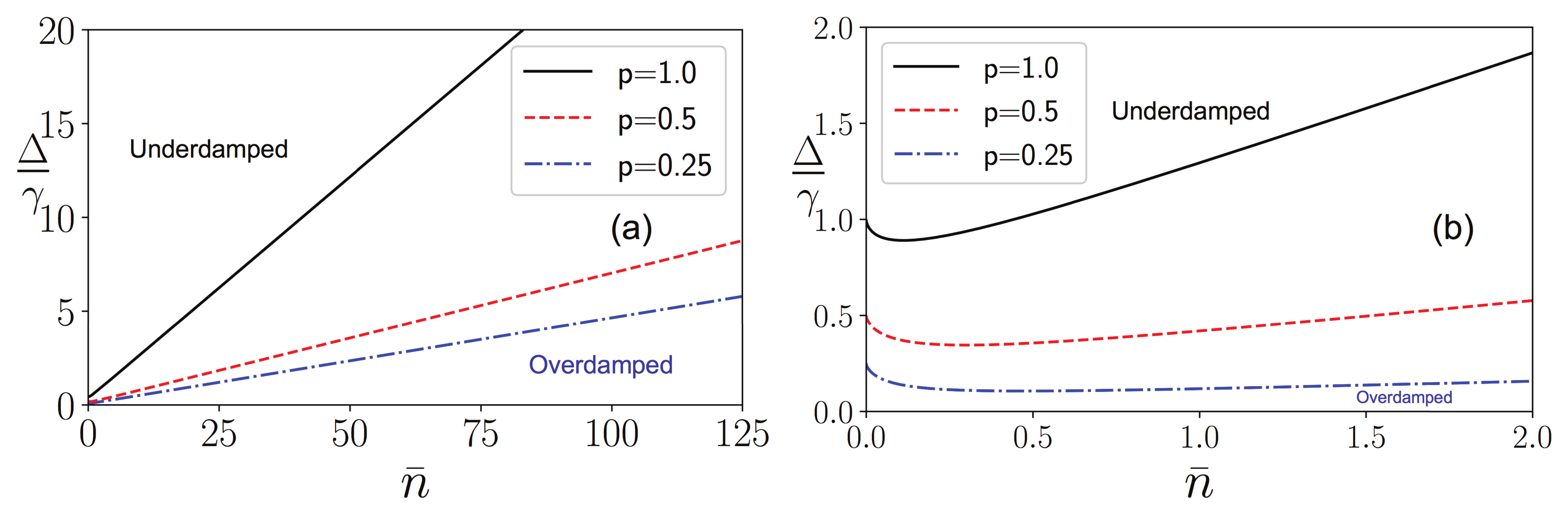}
     \caption{(a) Lines of zero discriminant separating the overdamped ($D<0$) from underdamped ($D>0$) regions for large ${\Delta}/{\gamma}$ and $\bar{n}$. 
     The dashed line corresponds to $\frac{\Delta}{\gamma}$ = $\bar{n}$. The overdamped behavior occurs in the upper left corner of the plot, where $\Delta/\gamma \gg \bar{n}$.   (b) A zoom into the small (${\Delta}/{\gamma}$, $\bar{n}$) region. Regions  below the zero-$D$ lines correspond to overdamped dynamics; those above the  zero-$D$ lines correspond to underdamped dynamics.  }\label{fig:Discriminant1D}
\end{figure}

\begin{figure}[h]
\captionsetup{singlelinecheck = false, format= hang, justification=raggedright, font=footnotesize, labelsep=space}
     \includegraphics[width=0.49\textwidth]{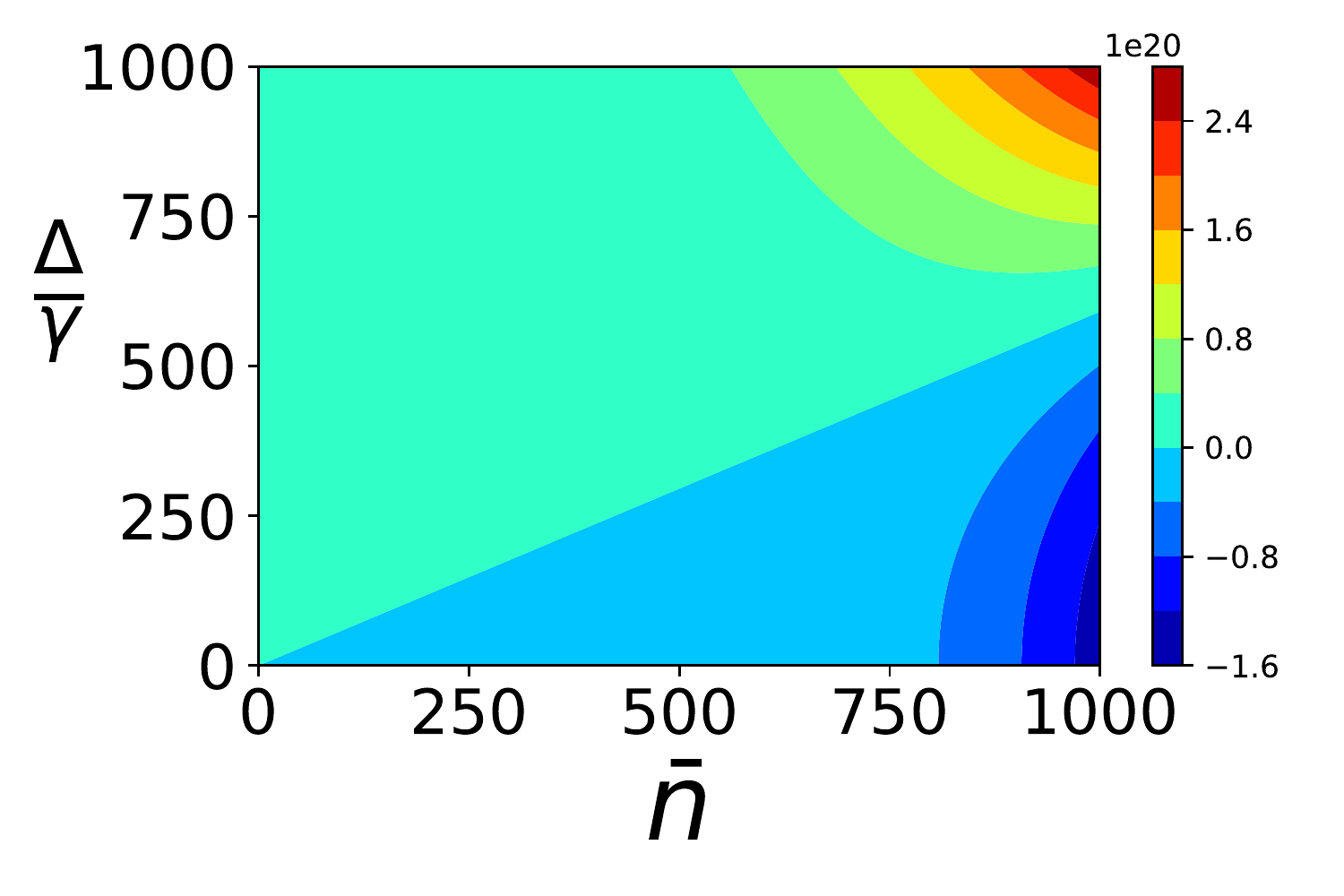} 
     \includegraphics[width=0.49\textwidth]{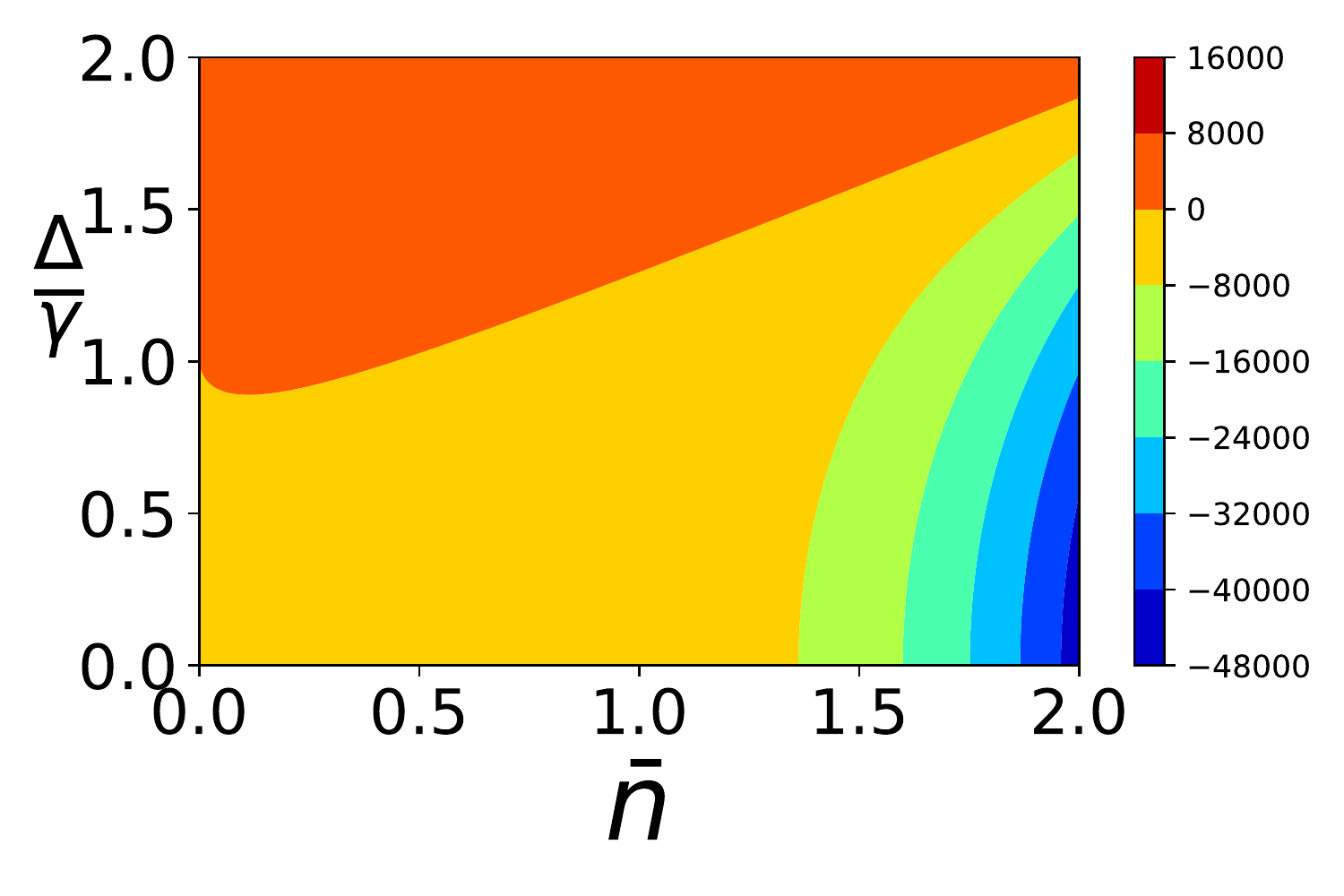}
     \caption{(a) A contour plot of the discriminant $D$ for large $\frac{\Delta}{\gamma}$ and $\bar{n}$ and $p = 1.0$; (b) A zoom into the region close to the origin ($\frac{\Delta}{\gamma}\ll1$ and $\bar{n}\ll1$ for $p=1.0$). Regions  of negative $D$ correspond to overdamped dynamics; those of positive $D$ correspond to underdamped dynamics.}\label{fig:Discriminant2D}
\end{figure}

\begin{figure}[t!]
\captionsetup{singlelinecheck = false, format= hang, justification=raggedright, font=footnotesize, labelsep=space}
\centering
\includegraphics[width=1.05\textwidth]{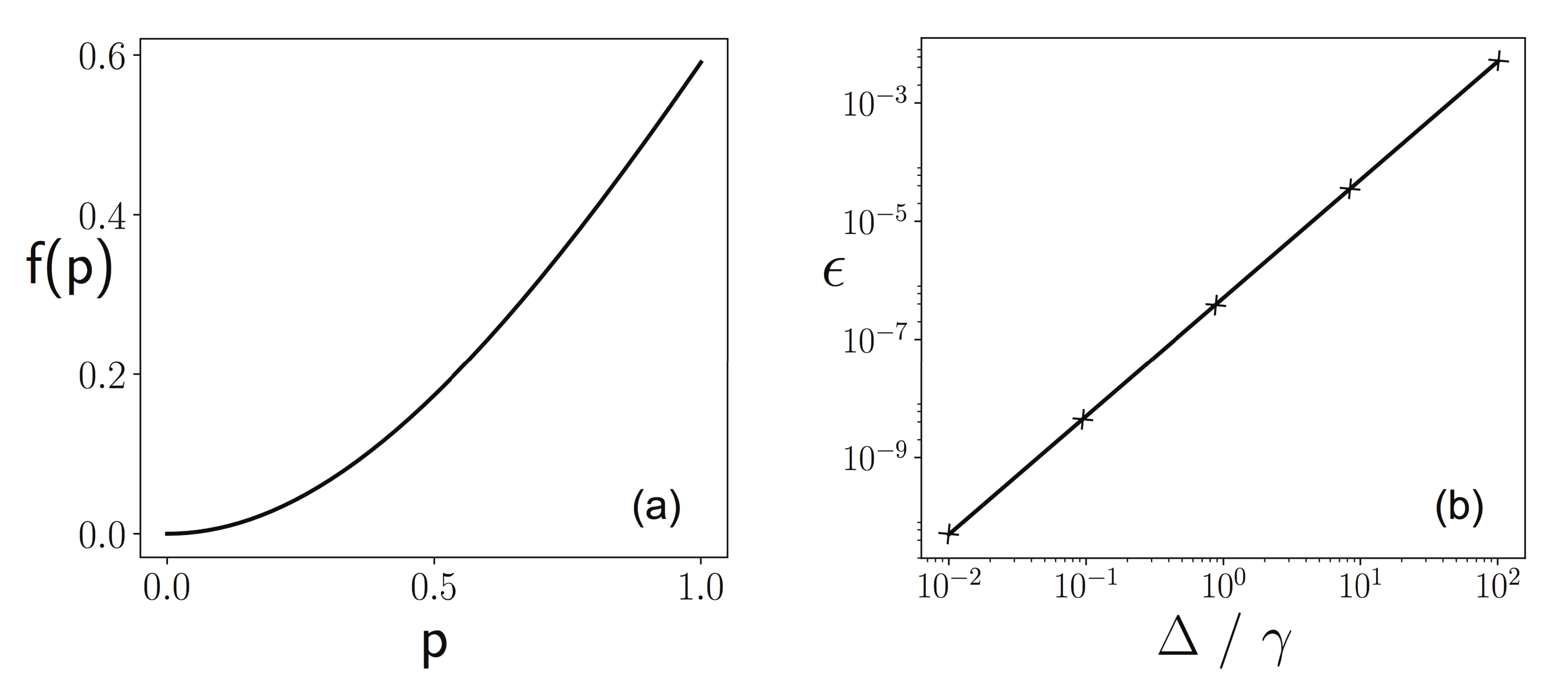}
\caption{(a) The universal function $f(p)$  that defines the boundary between the underdamped and overdamped regimes of the strongly driven V-system. (b)  $\epsilon = 1 - p_{c}$ as a function of $\Delta / \gamma$ for $\bar{n} = 10^{3}$. }
\label{fig:f_of_p}
\end{figure}


\begin{figure}[t]
\captionsetup{singlelinecheck = false, format= hang, justification=raggedright, font=footnotesize, labelsep=space}
\begin{center}
\includegraphics[width=1.05\textwidth]{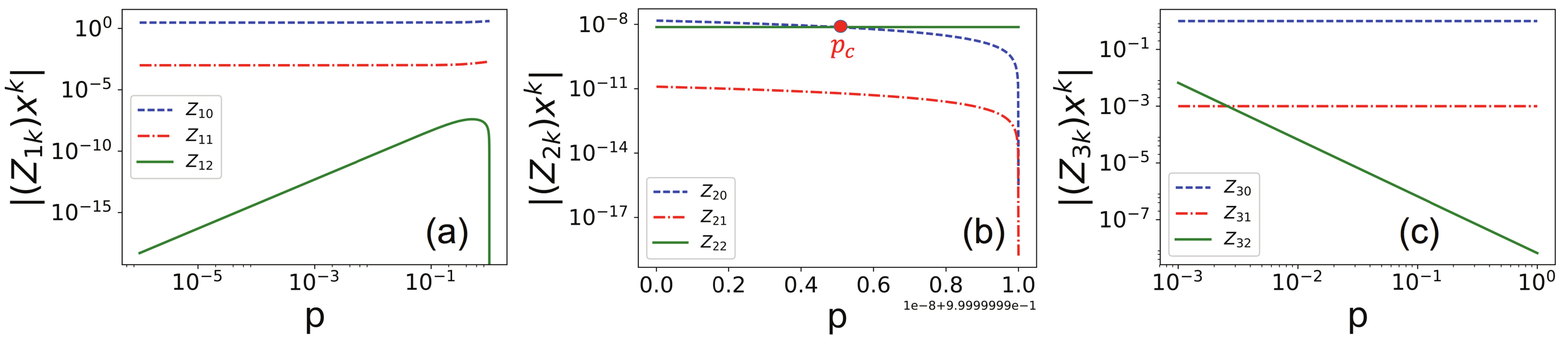}
\end{center}
\caption{Relative contributions of the different terms to the eigenvalues $\lambda_1$ (a), $\lambda_2$ (b) and $\lambda_3$ (c)  plotted as a function of  $p$ for $\Delta/\gamma=10^{-1}$, $\bar{n}=10^{3}$. The absolute value of $z_{jk} x^{k}$ are plotted because the $z_{jk}$ can take negative values.}
\label{fig:zjk_terms}
\end{figure}

\begin{figure}[h]
     \captionsetup{singlelinecheck = false, format= hang, justification=raggedright, font=footnotesize, labelsep=space}
     \includegraphics[width=1.05\textwidth]{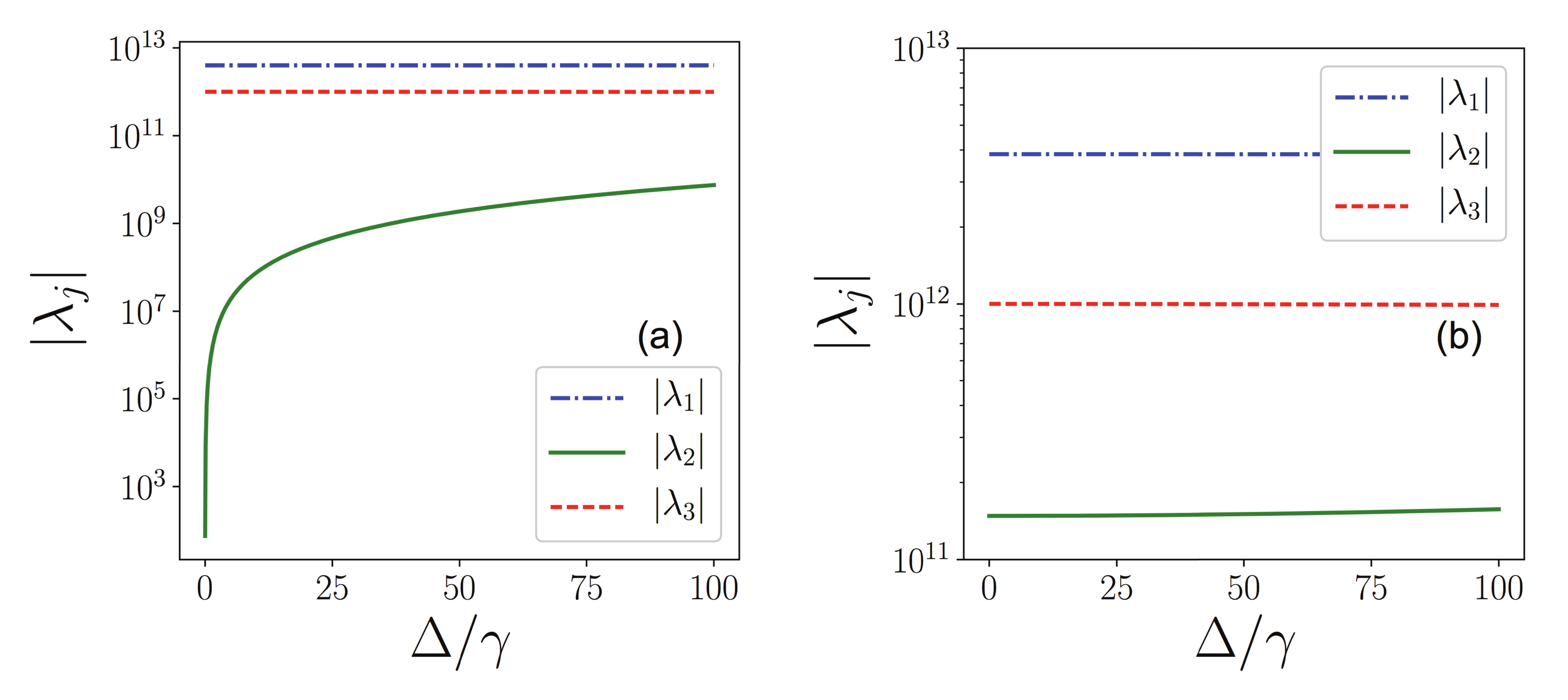} 
   \caption{(a) The eigenvalues  $\lambda_{j}$ ($j=1-3$)  of matrix $\mathbf{A}$ plotted vs ${\Delta}/{\gamma}$ for $\bar{n}$=$10^{3}$ and $p=1$ (b) Same as in panel (a) but for $\bar{n}$=$10^{3}$ and $p=0.9$. }
\label{fig:eigenvalues_vsDelta}
\end{figure}

\begin{figure}[t]
\captionsetup{singlelinecheck = false, format= hang, justification=raggedright, font=footnotesize, labelsep=space}
\begin{center}
\includegraphics[width=1.05\textwidth]{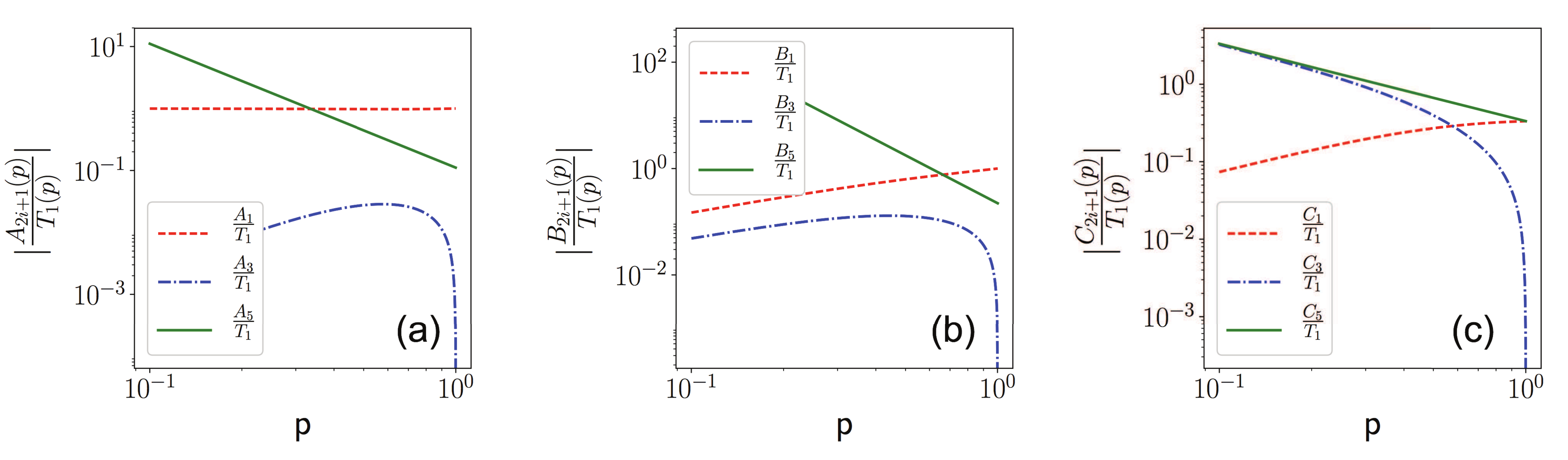}
\end{center}
\caption{Normalized contributions $A_{2i+1}/T_{1}(p)$, $B_{2i+1}/T_{1}(p)$, and $C_{2i+1}/T_{1}(p)$ ($i = 1-3$) to $\rho_{aa}(t)$ (a), $\rho^{R}_{ab}(t)$ (b) and $\rho^{I}_{ab}(t)$ (c) plotted as a function of  $p$ for $\frac{\Delta}{\gamma} << 1$.}
\label{fig:ABCodd_terms}
\end{figure}

\begin{figure}[t]
\captionsetup{singlelinecheck = false, format= hang, justification=raggedright, font=footnotesize, labelsep=space}
\begin{center}
\includegraphics[width=1.05\textwidth]{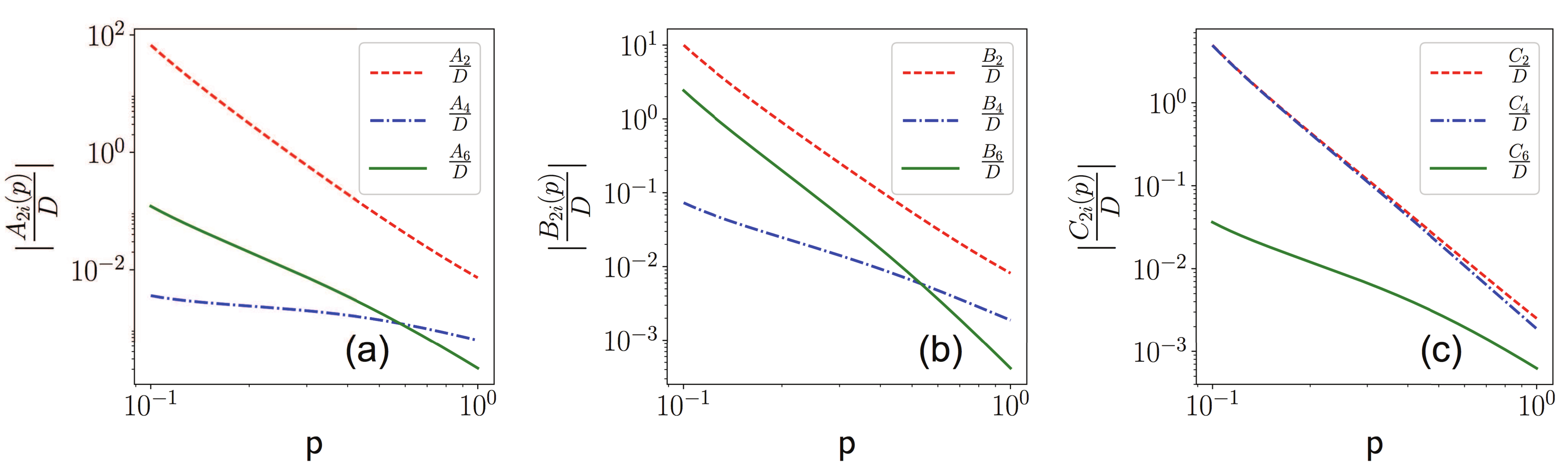}
\end{center}
\caption{Normalized contributions $A_{2i}$, $B_{2i}$, $C_{2i}$ ($i = 1-3$) to $\rho_{aa}(t)$ (a), $\rho^{R}_{ab}(t)$ (b) and $\rho^{I}_{ab}(t)$ (c)  plotted as a function of  $p$ for $\bar{n} = 10^{3}$, $\frac{\Delta}{\gamma} = 10^{2}$. The normalization factor $D = [T_{1}(p)+T_{2}(p) \frac{1}{\bar{n}^{2}} (\frac{\Delta}{\gamma})^{2} ]$.}
\label{fig:ABCeven_terms}
\end{figure}

\begin{figure}[h]
\captionsetup{singlelinecheck = false, format= hang, justification=raggedright, font=footnotesize, labelsep=space}
     \includegraphics[width=1.05\linewidth]{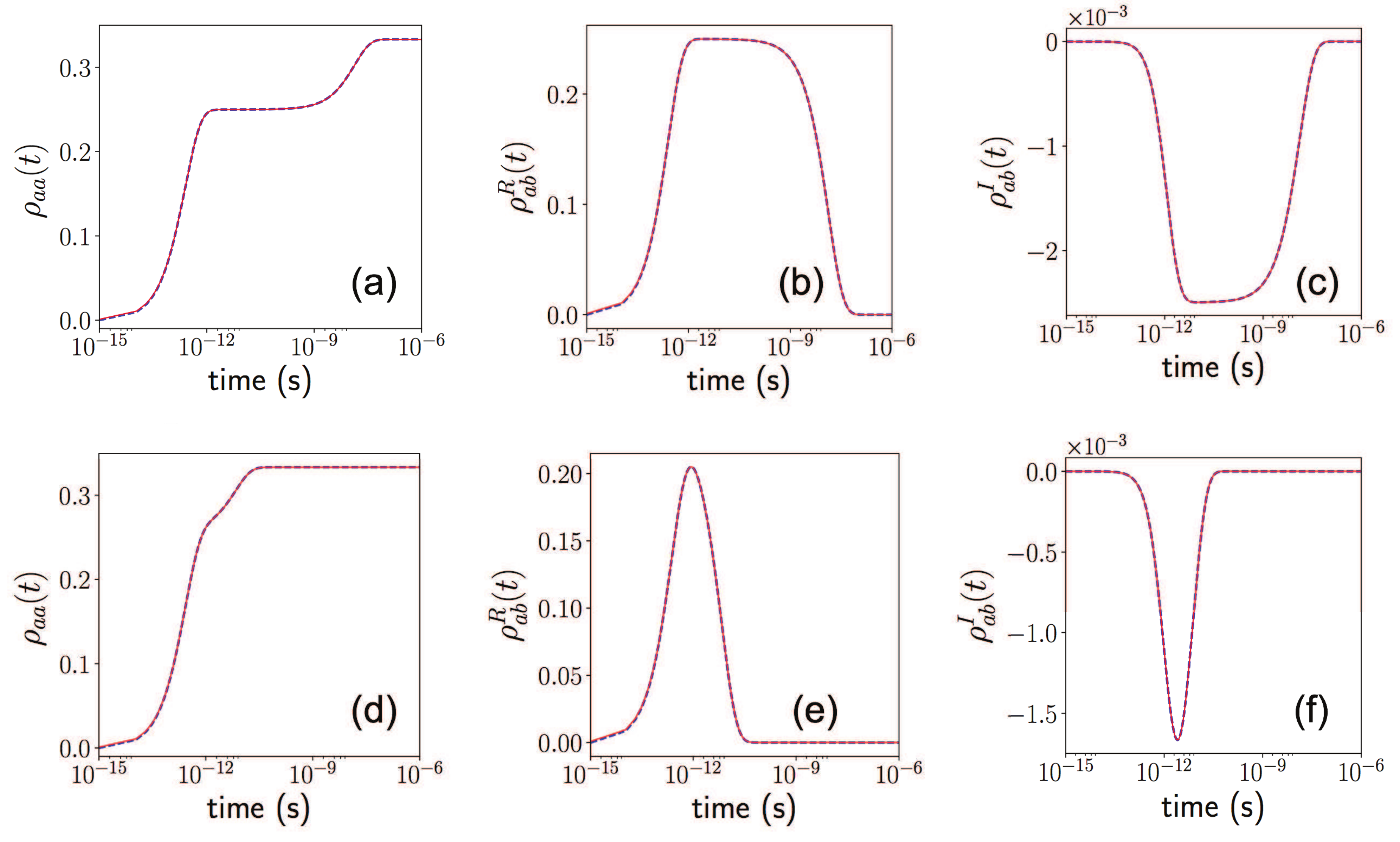} 
     \caption{Excited-state population [(a), (d)] and coherence [(b), (c), (e), (f)] dynamics of the symmetric V-system irradiated by incoherent light for $\bar{n}=10^{3}$ and $\frac{\Delta}{\gamma}=10$. The transition dipole alignment parameter $p$ is set to 1 [panels (a)-(c)] and to 0.9 [panels (c)-(d)]. Full lines -- analytical solution of the BR equations, dashed lines -- numerical expressions (\ref{GeneralSol1})-(\ref{GeneralSol3}) [panels (a)-(c)] and (\ref{GeneralSol1p})-(\ref{GeneralSol3p}) [panels (d)-(f)].}
\label{fig:dyn_largeDelta}     
\end{figure}

\begin{figure}[h]
\captionsetup{singlelinecheck = false, format= hang, justification=raggedright, font=footnotesize, labelsep=space}
     \includegraphics[width=1.05\linewidth]{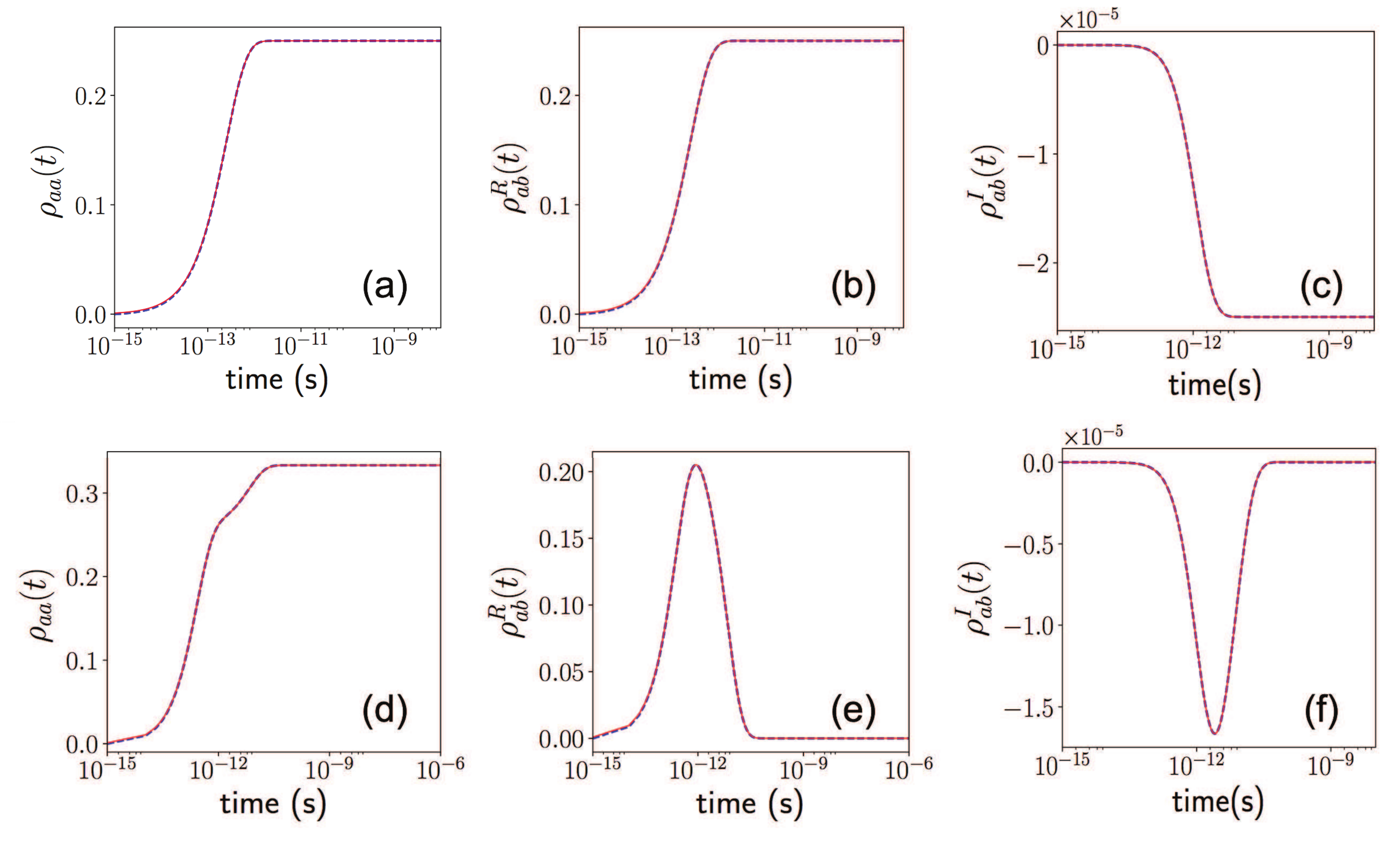} 
     \caption{Excited-state population [(a), (d)] and coherence [(b), (c), (e), (f)] dynamics of the symmetric V-system irradiated by incoherent light for $\bar{n}=10^{3}$ and $\frac{\Delta}{\gamma}=10^{-1}$. The transition dipole alignment parameter $p$ is set to 1 [panels (a)-(c)] and to 0.9 [panels (c)-(d)]. Full lines -- analytical solution of the BR equations, dashed lines -- numerical expressions (27) [panels (a)-(c)] and (29)-(31) [panels (d)-(f)]. The coherences in panel (c) will eventually decay to zero (not shown).}
\end{figure}
\label{fig:dyn_smallDelta}

\newpage

{\fontfamily{cmr}\selectfont \footnotesize 


\appendix*
\section{Analytic expressions}
  In this Appendix, we present a detailed derivation of the analytic expressions for the discriminant $\mathcal{D}$ and  the eigenvalues and eigenvectors of the coefficient matrix $\mathbf{A}$ given by Eq.~(4) of the main text as a function of the V-system parameters $\gamma$, $r$ and $\frac{\Delta}{\gamma}$. We also derive the analytic expressions for the populations and coherence dynamics which appear in Eqs. (18)-(20) and  (24)-(26) of the main text. The solutions are expressed in terms of the $p$-dependent coefficients listed in Tables 1-12.
  
  \subsection{Discriminant of the coefficient matrix A}
The general expression of the discriminant of the coefficient matrix is 
\begin{equation}
D= B^{3}+\big(C-\frac{3}{2}A\big(B+A^{2}\big)\big)^{2}
\end{equation}
where,
\begin{align}
A &=\frac{1}{3}\big(5r+3\gamma\big)\\
B &=\frac{\Delta^{2}}{3}-\frac{\gamma^{2}p^{2}}{3}+\gamma^{2}-\frac{4}{3}\gamma p^{2} r +\frac{10}{3}\gamma r -p^{2} r^{2}+\frac{7}{3} r^{2}- \frac{1}{9} \big(5r+3\gamma\big)^{2}\\
C &=\frac{1}{2}\Delta^{2} \gamma+\frac{3}{2}\Delta^{2}r-\frac{1}{2}\gamma^{3}p^{2}+\frac{1}{2}\gamma^{3}-\frac{5}{2}r\gamma^{2}-\frac{7}{2}\gamma p^{2} r^{2} + \frac{7}{2}\gamma r^{2}-\frac{3}{2} p^{2} r^{3}+\frac{3}{2} r^{3}+\frac{1}{27}\big(5r+3\gamma\big)^{3}
\end{align}
It is convenient to express the terms $A$, $B$, $C$ as a function of the occupation number $\bar{n}$
\begin{align}
A &= \frac{\gamma}{3}\big(3+5\bar{n}\big) \\
B &= \frac{\gamma^{2}}{3}\bigg[\frac{\Delta^{2}}{\gamma^{2}}-p^{2}-4p^{2}\bar{n}-\big(\frac{4}{3}+3p^{2}\big)\bar{n}^{2}\bigg] 
\end{align}
\begin{align}
\begin{split}
C &= \frac{\gamma^{3}}{2}\bigg[\frac{\Delta^{2}}{\gamma^{2}}+\big(3-p^{2}\big)+\big(15-5p^{2}+3\frac{\Delta^{2}}{\gamma^{2}}\big)\bar{n}+\big(\frac{71}{3}-7p^{2}\big)\bar{n}^{2}+\big(\frac{331}{27}-3p^{2}\big)\bar{n}^{3}\bigg] 
\end{split}
\end{align}
\begin{align}
C-\frac{3}{2}A(B+A^{2}) &= \frac{\gamma^{3}}{6}\bigg[ \big(4\frac{\Delta^{2}}{\gamma^{2}}+2p^{2}\big)\bar{n}+8p^{2}\bar{n}^{2}+\frac{(16+54p^{2})}{9}\bar{n}^{3}\bigg]
\end{align} 
Substituting Eqs. (A.6) and (A.8) into Eq. (A.1) 
we obtain the general expression of the discriminant as the polynomial function of the occupation number (i.e. $\bar{n}=\frac{r}{\gamma}$).
\begin{equation}
D=\frac{\gamma^{6}}{108}  \sum\limits_{k=0}^{6} d_{k} \bar{n}^{k},
\end{equation} 
where the expansion coefficients $d_{k}$ are listed in Table 1.

\subsubsection{Solving the equation $D=0$   in the strong pumping limit}
For large $\frac{\Delta}{\gamma}$ and $\bar{n}$, the significant terms in Eq. (9) are the 6th order terms  $d_{0}$, $d_{4}\bar{n}^{4}$ and $d_{6} \bar{n}^{6}$ \\
\begin{equation}
D=\frac{\gamma^{6}}{108}\big(d_{0} + d_{4} \bar{n}^{4} + d_{6} \bar{n}^{6}\big).
\end{equation}
To solve  the equation $D = 0$, we take $\frac{\Delta}{\gamma} = y$, $\bar{n} = x$ and simplify as
\begin{align}
4\big(y^{2}-p^{2}\big)^{3} + \bigg[\frac{8}{3}\big(8+27 p^{2}\big)\big(2y^{2}+p^{2}\big)+\frac{4}{3}\big(4+9p^{2}\big)^{2}\big(y^{2}-p^{2}\big)-64p^{4}\big(1+9p^{2}\big)\bigg] x^{4} - 36p^{4}\big(1+3p^{2}\big) x^{6} &= 0 \nonumber \\
4y^{6} + \bigg[\frac{8}{3}\big(8+27 p^{2}\big) 2y^{2}+\frac{4}{3}\big(4+9p^{2}\big)^{2} y^{2} - 64p^{4}\big(1+9p^{2}\big)\bigg] x^{4} - 36p^{4}\big(1+3p^{2}\big) x^{6} &= 0 \nonumber 
\end{align}
Dividing on both sides by $x^{6}$, we get
\begin{equation}
4\bigg(\frac{y}{x}\bigg)^{6} + \bigg[\frac{16}{3}\big(8+27 p^{2}\big) + \frac{4}{3}\big(4+9p^{2}\big)^{2}\bigg] \bigg(\frac{y}{x}\bigg)^{2} - 64p^{4}\big(1+9p^{2}\big) \frac{1}{x^{2}} - 36p^{4}\big(1+3p^{2}\big) = 0 \nonumber 
\end{equation}
Neglecting the term proportional to $\frac{1}{x^{2}}$ for large $\bar{n} = x$ and defining $(\frac{y}{x})^{2} = z$, the above equation reduces to the form of a depressed cubic 
\begin{align} 
z^{3} + P z + Q &= 0,
\end{align} 
where
\begin{align}
P &= \big(16+ 60 p^{2} + 27p^{4}\big) \\ 
Q &= - 9p^{4}\big(1+3p^{2}\big)
\end{align} 
To solve the depressed cubic equation, we substitute $z = u + v$ into Eq. (11)
which becomes 
\begin{equation}
u^{3} + v^{3} + \big(u+v\big) \big(3uv+P\big) + Q = 0 
\end{equation} 
The arbitary variables $u$, $v$ are choosen in such a way that 
\begin{align}
\big(3uv+P\big) &= 0 \nonumber \\
uv = - \frac{P}{3} 
\end{align} 
Cubing Eq. (15) on both sides and expressing $v^{3}$ in terms of $u^{3}$, we get
\begin{equation}
v^{3} = -\frac{P^{3}}{27} \frac{1}{u^{3}}
\end{equation} 
Using Eqs. (A.15) and (A.16) into Eq. (A.14), we rearrange terms to get a quadratic equation in $u^{3} = t$ 
\begin{equation}
t^{2} + Q t - \frac{4P^{3}}{27} = 0
\end{equation} 
The two roots of the above equation are 
\begin{equation}
t_{1} = -\frac{Q}{2} + \sqrt{\frac{Q^{2}}{4}+\frac{P^{3}}{27}}
\end{equation}
\begin{equation}
t_{2} = -\frac{Q}{2} - \sqrt{\frac{Q^{2}}{4}+\frac{P^{3}}{27}}
\end{equation}  
We set $u^{3} = t_{1}$, $v^{3} = t_{2}$ which satisfy the required conditions $u^{3}+v^{3} = -Q$, $u^{3} v^{3} = -\frac{P^{3}}{27}$. This shows that $u = \sqrt[3]{t_{1}}$ and $v = \sqrt[3]{t_{2}}$ are solutions to Eq. (A.14). As $z = (\frac{y}{x})^{2}$ can not be complex valued and $\sqrt[3]{t_{1}}$, $\sqrt[3]{t_{1}}$ are real and positive for all $p$ values, the real solution for $z$ is 
\begin{align}
z &= u + v \nonumber \\
\bigg(\frac{y}{x}\bigg)^{2} &= \sqrt[3]{t_{1}} + \sqrt[3]{t_{2}} \nonumber \\ 
\frac{\Delta}{\gamma} &= \sqrt{\sqrt[3]{t_{1}} + \sqrt[3]{t_{2}}} \enspace \bar{n} \nonumber \\
\frac{\Delta}{\gamma} &= f(p) \enspace \bar{n}
\end{align} 
where 
\begin{equation}
f(p) = \sqrt{\sqrt[3]{t_{1}} + \sqrt[3]{t_{2}}}
\end{equation} 
with $t_{1}$ and $t_{2}$ given by Eqs. (A.18) and (A.19). 
This shows that in the strong pumping limit and when $\frac{\Delta}{\gamma}$ and $\bar{n}$  are large, the critical $D=0$ line behaves as a straight line with slope given by $m = f(p)$, a function of $p$ only. A plot of $f(p)$ is shown in Fig. 4(a) of the main text.

\subsection{Eigenvalues of matrix A}
The eigenvalues $\lambda_k$ of the coefficient matrix $\mathbf{A}$ are given by Cardano's solution of the characteristic equation 
\begin{equation}
\lambda_{j} = -A + 
 \alpha_{j} \frac{B}{\mathcal{T}} -
\beta_{j} \mathcal{T} \enspace (j=1-3),
\end{equation}\\
where
\begin{align}
\mathcal{T} &= \sqrt[3]{E+\sqrt{D}} \\
E &= (C-\frac{3}{2}A(B+A^{2}))\\
D &= B^{3}+(C-\frac{3}{2}A(B+A^{2}))^{2} \\
\omega &=\frac{-1+i\sqrt{3}}{2}\\
\omega^{2} &=\frac{-1-i\sqrt{3}}{2}
\end{align}
and $(\alpha_{1}, \beta_{1}) = (1, 1), \enspace (\alpha_{2}, \beta_{2}) = (\omega^{2}, \omega), \enspace (\alpha_{3}, \beta_{3}) = (\omega, \omega^{2})$.

\subsubsection{Eigenvalues in the overdamped regime [$\Delta / (\bar{n} \gamma) < f(p)$]}
In the strong pumping limit where $\bar{n} \gg 1$, we define a new variable $x = 1 / \bar{n} \ll 1$ and express the terms $\mathcal{D}$ and $E$ in the polynomial form of $x= 1 / \bar{n}= \gamma / r$. We find the expression for $E$ by rearranging Eq. (A.8) as 
\begin{equation}
E = \frac{r^{3}}{6}\sum\limits_{k=1}^{3} c_{k} x^{3-k}
\end{equation}
where the $p$-dependent expansion coefficients $c_{k}$ ($k$ = 1, 2, 3) are listed in Table 1. \\
In order to simplify the term $\mathcal{T}$ in the eigenvalue expression, we first express $\sqrt{D}$ [with  $D$  given by Eq. (A.9)] in the following form
\begin{equation}
\sqrt{D}=r^{3}\sqrt{\frac{d_{6}}{108}}\sqrt{(1+\alpha(x))}
\end{equation} 
where,
\begin{equation}
\alpha(x) = \frac{1}{d_{6}} \sum\limits_{k=1}^{6} d_{6-k} x^{k}
\end{equation} 
In the strong pumping limit \big($x= 1 / \bar{n} \ll 1$\big)  the terms $\frac{d_{6-k}}{d_{6}} x^{k}$, $k \geq 1$ are all negligible compared to 1 and thus $ |\alpha(x)| \ll 1$. The binomial expansion presented here and all the succeeding expansions are valid for $\bar{n} \ge 10^{2}$ and $p > 0.1$ \big(for $\Delta/\gamma<1$\big) and $0.89 < p < 1$ \big(for $\Delta/\gamma>1$\big). Taking the binomial expansion of $\sqrt{1+\alpha(x)}$. 
\begin{equation}
(1+\alpha)^{\frac{1}{2}}=1+\frac{1}{2}\alpha-\frac{1}{8}\alpha^{2}+\frac{3}{48}\alpha^{3}-\frac{15}{384}\alpha^{4}+\frac{105}{3840}\alpha^{5}- ......
\end{equation} 
we find the terms  $\alpha^k$ with $k\le 4$ by using a multinomial expansion
\begin{align}
\alpha^{2} &=(\frac{d_{5}}{d_{6}})^{2}x^{2} + (2\frac{d_{4}d_{5}}{d_{6}^{2}})x^{3}+({2\frac{d_{3}d_{5}}{d_{6}^{2}}+(\frac{d_{4}}{d_{6}})^{2}})x^{4}+(2\frac{d_{2}d_{5}}{d_{6}^{2}}+2\frac{d_{3}d_{4}}{d_{6}^{2}})x^{5}+(2\frac{d_{1}d_{5}}{d_{6}^{2}}+2\frac{d_{2}d_{4}}{d_{6}^{2}}+(\frac{d_{4}}{d_{6}})^{3})x^{6} \nonumber \\
\alpha^{3} &=(\frac{d_{5}}{d_{6}})^{3}x^{3}+(3\frac{d_{4}d_{5}^{2}}{d_{6}^{3}})x^{4}+(3\frac{d_{3}d_{5}^{2}}{d_{6}^{3}}+3\frac{d_{4}^{2}d_{5}}{d_{6}^{3}})x^{5}+(3\frac{d_{2}d_{5}^{2}}{d_{6}^{3}}+6\frac{d_{3}d_{4}d_{5}}{d_{6}^{3}}+(\frac{d_{4}}{d_{6}})^{3})x^{6} \nonumber \\
\alpha^{4} &=(\frac{d_{5}}{d_{6}})^{4}x^{4}+(4\frac{d_{4}d_{5}^{3}}{d_{6}^{4}})x^{5}+(4\frac{d_{3}d_{5}^{3}}{d_{6}^{4}}+6\frac{d_{4}^{2}d_{5}^{2}}{d_{6}^{4}})x^{6} \nonumber
\end{align} 
Substituting Eq. (A.30) and the above expressions into Eq. (A.29), we get
\begin{equation}
\sqrt{D}=r^{3}\sqrt{\frac{d_{6}}{108}}  \Big[1 + \sum\limits_{k=1}^{6} u_{k} x^{k}\Big]
\end{equation} 
where the expansion coefficients $u_{k}$ are listed in Table 2. 
Now, using Eqs. (A.28), (A.32) in Eq. (A.23) the term $\mathcal{T}$ can be expanded as 
\begin{align}
\mathcal{T} &= \Big[\frac{r^{3}}{6} \sum\limits_{k=1}^{3} c_{k} {x}^{3-k}+ r^{3}\sqrt{\frac{d_{6}}{108}}(1+ \sum\limits_{k=1}^{6} u_{k} {x}^{k})\Big]^{\frac{1}{3}} \nonumber \\
&= r \sqrt[3]{\frac{c_{3}}{6}+\sqrt{\frac{d_{6}}{108}}}\Bigg[1+\frac{ \sum\limits_{k=1}^{2}(\frac{c_{3-k}}{6}+\sqrt{\frac{d_{6}}{108}}u_{k})}{\sqrt[3]{\frac{c_{3}}{6}+\sqrt{\frac{d_{6}}{108}}}}x^{k}+\frac{ \sum\limits_{k=3}^{6}\sqrt{\frac{d_{6}}{108}}u_{k}}{\sqrt[3]{\frac{c_{3}}{6}+\sqrt{\frac{d_{6}}{108}}}}x^{k}\Bigg]^{\frac{1}{3}} \nonumber \\
&=r K \Big[1+ \sum\limits_{k=0}^{6} b_{k} x^{k}\Big]^{\frac{1}{3}} 
\end{align}
where the term $K$ and the coefficients $b_{k}$ are listed in Table 1. 

Equation (A.33) can be rewritten as
\begin{equation}
\mathcal{T} = rK \Big[1+\beta(x)\Big]^{\frac{1}{3}},
\end{equation}
where
\begin{equation}
\beta(x) =  \sum\limits_{k=1}^{6} b_{k} x^{k} \nonumber
\end{equation} 
In the strong pumping limit \big($x= 1 / \bar{n} \ll 1$\big),  the terms $b_{k} x^{k}$, $k \geq 1$ are all negligible compared to 1 and thus $ |\beta(x)| \ll 1$. Taking the binomial expansion of $\sqrt[3]{1+\beta}$. 
\begin{equation}
(1+\beta)^{\frac{1}{3}}=1+\frac{1}{3}\beta-\frac{1}{9}\beta^{2}+\frac{5}{81}\beta^{3}-\frac{10}{243}\beta^{4}+ ...,
\end{equation} 
 evaluating the  terms  $\beta^k$ with $k\le 3$ using the multinomial expansion
\begin{align}
\beta^{2} &= b_{1}^{2}x^{2}+2b_{1}b_{2}x^{3}+(2b_{1}b_{3}+b_{2}^{2})x^{4}+(2b_{1}b_{4}+2b_{2}b_{3})x^{5}+(2b_{1}b_{5}+2b_{2}b_{4}+b_{3}^{2})x^{6}+... \nonumber \\
\beta^{3} &= b_{1}^{3}x^{3}+(3b_{1}^{2}b_{2})x^{4}+(3b_{1}^{2}b_{3}+3b_{1}b_{2}^{2})x^{5}+(3b_{1}^{2}b_{4}+6b_{1}b_{2}b_{3}+b_{2}^{3})x^{6}+ .... \nonumber 
\end{align} 
and substituting the result in Eq. (A.34), we get
\begin{equation}
\mathcal{T} = rK \Big[1+ \sum\limits_{k=0}^{6} v_{k} x^{k}\Big]
\end{equation} 
where the expansion coefficients $v_{k}$ are listed in Table 2. 
The second term in the eigenvalue expression contains the fraction $\frac{1}{\mathcal{T}}$, which  we simplify to obtain 
\begin{align}
\frac{1}{\mathcal{T}} &= \frac{1}{r K}\Big[1+ \sum\limits_{k=0}^{6} v_{k} x^{k}\Big]^{-1} \nonumber \\
&=\frac{1}{r K}\Big[1+\Lambda(x)\Big]^{-1}
\end{align}
where,
\begin{equation}
\Lambda(x) =  \sum\limits_{k=0}^{6} v_{k} x^{k}
\end{equation} \\
In the strong pumping limit \big($x=1 / \bar{n} <<1$\big)  the terms $v_{k} x^{k}$, $k \geq 1$ are all negligible compared to 1 and thus $ |\Lambda(x)| << 1$. Taking the binomial expansion of $(1+\Lambda)^{-1}$
\begin{equation}
(1+\Lambda)^{-1}=1-\Lambda+\Lambda^{2}-\Lambda^{3}+\Lambda^{4}-\Lambda^{5}+ .....
\end{equation}
we find the  terms $\Lambda^k$ with $k\le 4$ by using the multinomial expansion
\begin{align} 
\Lambda^{2} &= v_{1}^{2}x^{2}+2v_{1}v_{2}x^{3}+(2v_{1}v_{3}+v_{2}^{2})x^{4}+(2v_{1}v_{4}+2v_{2}v_{3})x^{5}+(2v_{1}v_{5}+2v_{2}v_{4}+v_{3}^{2})x^{6} \nonumber \\
\Lambda^{3} &= v_{1}^{3}x^{3}+3v_{1}^{2}v_{2}x^{4}+(3v_{1}^{2}v_{3}+3v_{1}v_{2}^{2})x^{5}+(3v_{1}^{2}v_{4}+6v_{1}v_{2}v_{3}+v_{3}^{2})x^{6} \nonumber \\
\Lambda^{4} &= v_{1}^{4}x^{4}+4v_{1}^{3}v_{2}x^{5}+(4v_{1}^{3}v_{3}+6v_{1}^{2}v_{2}^{2})x^{6} \nonumber 
\end{align} 
Substituting the above expressions into Eq. (A.37), we obtain
\begin{equation}
\frac{1}{\mathcal{T}}=\frac{1}{r K}\Big[1+ \sum\limits_{k=0}^{6} \mathcal{W}_{k} x^{k}\Big],
\end{equation}
where the expansion coefficients $\mathcal{W}_{k}$ are listed in Table 3.  
Now we evaluate the second term $\frac{B}{\mathcal{T}}$ of the eigenvalue expression given by Eq. (A.22). In the polynomial form of $x= 1 / \bar{n}$, we have
\begin{equation}
B=\frac{r^{2}}{3}\Big[\Big(\frac{\Delta^{2}}{\gamma^{2}}-p^{2}\Big)x^{2}-4p^{2}x-\Big(\frac{4}{3}+3p^{2}\Big)\Big]
\end{equation} \\
Multiplying Eq. (A.40) and Eq. (A.41), we get 
\begin{equation}
\frac{B}{\mathcal{T}}=\frac{r^{2}}{3}\Big[\Big(\frac{\Delta^{2}}{\gamma^{2}}-p^{2}\Big)x^{2}-4p^{2}x-\Big(\frac{4}{3}+3p^{2}\Big)] \times \frac{1}{r K}\Big[1+ \sum\limits_{k=0}^{6} \mathcal{W}_{k} x^{k}\Big] \nonumber 
\end{equation} 
\begin{equation}
\begin{split}
= \frac{r}{3K}\Big[-\Big(\frac{4}{3}+3p^{2}\Big)-\Big[\Big(\frac{4}{3}+3p^{2}\Big)\mathcal{W}_{1}+4p^{2}\Big]x+\Big[\Big(\frac{\Delta^{2}}{\gamma^{2}}-p^{2}\Big)-4p^{2}\mathcal{W}_{1}-\Big(\frac{4}{3}+3p^{2}\Big)\mathcal{W}_{2}\Big]x^{2}\\+\Big[\Big(\frac{\Delta^{2}}{\gamma^{2}}-p^{2}\Big)\mathcal{W}_{1}-4p^{2}\mathcal{W}_{2}-\Big(\frac{4}{3}+3p^{2}\Big)\mathcal{W}_{3}\Big]x^{3}+\Big[\Big(\frac{\Delta^{2}}{\gamma^{2}}-p^{2}\Big)\mathcal{W}_{2}-4p^{2}\mathcal{W}_{3}-\Big(\frac{4}{3}+3p^{2}\Big)\mathcal{W}_{4}\Big]x^{4}\\+\Big[\Big(\frac{\Delta^{2}}{\gamma^{2}}-p^{2}\Big)\mathcal{W}_{3}-4p^{2}\mathcal{W}_{4}-\Big(\frac{4}{3}+3p^{2}\Big)\mathcal{W}_{5}\Big]x^{5}+\Big[\Big(\frac{\Delta^{2}}{\gamma^{2}}-p^{2}\Big)\mathcal{W}_{4}-4p^{2}\mathcal{W}_{5}-\Big(\frac{4}{3}+3p^{2}\Big)\mathcal{W}_{6}\Big]x^{6}\\+\Big[\Big(\frac{\Delta^{2}}{\gamma^{2}}-p^{2}\Big)\mathcal{W}_{5}-4p^{2}\mathcal{W}_{6}\Big]x^{7}+\Big(\frac{\Delta^{2}}{\gamma^{2}}-p^{2}\Big)\mathcal{W}_{6}\Big]x^{8} 
\end{split}
\end{equation}
Substituting the expressions of $A$, $\frac{B}{\mathcal{T}}$ and $\mathcal{T}$ in Eq. (A.22) and simplifying, we get Eq. (A.9) of the main text for the eigenvalues of the coefficient matrix 
\begin{equation}
\lambda_{j}=r\sum_{k=0}^{8}z_{jk} x^{k} \enspace  (j = 1 - 3),
\end{equation}
where the expansion coefficients $z_{jk}$ are listed in Table 3.

\subsection{Eigenvectors of matrix $\mathbf{A}$}
The general expression for the eigenvectors of the coefficient matrix $\mathbf{A}$ is obtained by solving the system of linear equations $\big(\mathbf{A} - \lambda_{j}\big) V_{j} = 0$ to yield 
\begin{equation}
V_{j} =
\begin{bmatrix}
       \frac{\Delta p(r+\gamma)}{\mathcal{D}_{j}}  \\
       -\frac{\Delta (3r+\gamma+\lambda_{j})}{\mathcal{D}_{j}} \\
       1.0  \\
\end{bmatrix} \enspace  (j = 1 - 3),
\end{equation}
where
\begin{equation}
\mathcal{D}_{j}=-\lambda_{j}^{2}-2(\gamma+2r)\lambda_{j}-(1-p^{2})\gamma^{2}-4\gamma r(1-p^{2})+3r^{2}(1-p^{2}).
\end{equation}

\subsubsection{Eigenvectors in the overdamped regime [$\Delta / (\bar{n} \gamma) < f(p)$]}
In the strong pumping regime ($x\ll 1$),  the terms $x^{n}$ with $n \geq 3$ in Eq. (A.43) can be neglected and the eigenvalues are given by
\begin{equation}
\lambda_{j}=r\Big[z_{j0}+z_{j1}x+z_{j2}x^{2}\Big]
\end{equation}
To evaluate the term $\mathcal{D}_{j}$, we evaluate the square of $\lambda_{j}$ as 
\begin{equation}
\lambda_{j}^{2} = r^{2}\Big[z_{j0}^{2}+2z_{j0}z_{j1}x+\Big(z_{j1}^{2}+2z_{j0}z_{j2}\Big)x^{2}+2z_{j1}z_{j2}x^{3}+z_{j2}^{2}x^{4}\Big]
\end{equation} 
Substituting Eqs. (A.46), (A.47) in Eq. (A.45), we obtain
\begin{equation}
\mathcal{D}_{j} = -r^{2} \sum_{k=0}^{4}L_{jk} x^{k}
\end{equation}
where the expansion coefficients $L_{jk}$ are listed in Table 6. 
For the first eigenvector $\vec{V}_{1}$, we find
\begin{equation}
\mathcal{D}_{1} = -r^{2} \sum_{k=0}^{4}L_{1k} x^{k}
\end{equation} 
In particular, the terms $L_{10}$, $L_{11} x$ are negligible compared to other terms in Eq. (A.49) and are dropped. 
To find $\frac{1}{\mathcal{D}_{1}}$ required to evaluate Eq. (A.44), we proceed as follows
\begin{align}
\frac{1}{\mathcal{D}_{1}} &=-\frac{1}{r^{2}L_{12}x^{2}\Big[1+\frac{L_{13}}{L_{12}}x+\frac{L_{14}}{L_{12}}x^{2}\Big]} \nonumber \\
&=-\frac{1}{r^{2}L_{12}x^{2}} \Big[1+\alpha(x)\Big]^{-1}
\end{align} 
where, 
\begin{equation}
\alpha(x) = \frac{L_{13}}{L_{12}}x+\frac{L_{14}}{L_{12}}x^{2}
\end{equation}
For $x \ll 1$,\enspace $\alpha(x)=\Big(\frac{L_{13}}{L_{12}}x+\frac{L_{14}}{L_{12}}x^{2}\Big) \ll 1$ and we can use the binomial expansion to get
\begin{equation}
(1+\alpha)^{-1}=1-\alpha+\alpha^{2}-\alpha^{3}+ \ldots
\end{equation} 
Evaluating the terms up to the third order in $\alpha$ 
\begin{align}
\alpha^{2} &=\Big(\frac{L_{13}}{L_{12}}\Big)^{2}x^{2}+\frac{2L_{13}L_{14}}{L_{12}^{2}}x^{3}+\Big(\frac{L_{14}}{L_{12}}\Big)^{2}x^{4} \\
\alpha^{3} &=\Big(\frac{L_{13}}{L_{12}}\Big)^{3}x^{3}+\frac{3L_{13}^{2}L_{14}}{L_{12}^{3}}x^{4}+\frac{3L_{13}L_{14}^{2}}{L_{12}^{3}}x^{5}+\Big(\frac{L_{14}}{L_{12}}\Big)^{3}x^{6}
\end{align}
and using Eqs. (A.51) to (A.54) into Eq. (A.50), we get 
\begin{equation}
\frac{1}{\mathcal{D}_{1}}=-\frac{1}{r^{2}L_{12}x^{2}} \Big[1+\sum_{m=1}^{6} k_{1m} x^{m}\Big]
\end{equation}
where the expansion coefficients $k_{1m}$ are listed in Table 6. \\ 
We can now evaluate the first component of the eigenvector $\vec{V_{1}}$ as follows
\begin{align}
V_{11} &=\frac{\Delta p (\gamma+r)}{\mathcal{D}_{1}} \nonumber \\
&=-\frac{p}{L_{12}}\frac{\Delta r(1+\frac{\gamma}{r})}{r^{2}(\frac{\gamma}{r})^{2}}\Big[1+\sum_{m=1}^{6} k_{1m} x^{m}\Big] \nonumber \\
&=-\frac{p}{L_{12}}\Big(\frac{\Delta}{\gamma}\Big)\bar{n} \sum_{m=0}^{7}a_{1m}x^{m}
\end{align}
where the expansion coefficients $a_{1m}$ are listed in Table 6. 
Proceeding in a similar way, we find the second component of the eigenvector $\vec{V}_{1}$ as
\begin{equation}
V_{12}=\frac{1}{L_{12}}\Big(\frac{\Delta}{\gamma}\Big)\bar{n}\sum_{m=0}^{8}b_{1m}x^{m}
\end{equation}
where the expansion coefficients $b_{1m}$ are listed in Table 6. The third component of the eigenvector $\vec{V}_{1}$ is $V_{13}=1$. \\
Combining the expressions for $V_{11}$, $V_{12}$ and $V_{13}$, we obtain the first eigenvector as
\begin{equation}
\vec{V}_{1} = 
\begin{bmatrix}
     -\frac{p}{L_{12}}\Big(\frac{\Delta}{\gamma}\Big)\bar{n} \sum_{m=0}^{7}a_{1m}x^{m}\\
    \frac{1}{L_{12}}\Big(\frac{\Delta}{\gamma}\Big)\bar{n}\sum_{m=0}^{8}b_{1m}x^{m}  \\
       1 \\
\end{bmatrix}
\end{equation} 
Proceeding in a similar way as for the first eigenvector $\vec{V}_{1}$, we evaluate $\mathcal{D}_{2}$ and the components of the second eigenvector $\vec{V}_{2}$
\begin{equation}
\vec{V}_{2} = 
\begin{bmatrix}
     -\frac{p}{L_{22}}\Big(\frac{\Delta}{\gamma}\Big)\bar{n} \sum_{m=0}^{7}a_{2m}x^{m}\\
    \frac{1}{L_{22}}\Big(\frac{\Delta}{\gamma}\Big)\bar{n}\sum_{m=0}^{8}b_{2m}x^{m}  \\
       1 \\
\end{bmatrix},
\end{equation} 
where the coefficients $L_{2k}$, $k_{2m}$, $a_{2m}$, $b_{2m}$ are listed in Table 6 and $z_{2k}$  ($k=0- 2$) are evaluated with $\alpha_{2}=\enspace \omega^{2},\enspace \beta_{2}=\enspace \omega$. \\
For the third eigenvector $\vec{V}_{3}$, the term $\mathcal{D}_{3}$ is given by
\begin{equation}
\mathcal{D}_{3} = -r^{2} \sum_{k=0}^{4} L_{3k} x^{k}
\end{equation}
where the coefficients $L_{3k}$ are listed in Table 6 and $z_{3k}$ ($k=0- 2$) are evaluated with $\alpha_{3}=\enspace \omega,\enspace \beta_{3}=\enspace \omega^{2}$. \\
Unlike in the case of the first and second eigenvectors, the terms $L_{30}$ and $L_{31}x$ are not negligible compared to other $L_{3k} x^{k}$ terms. To find $\frac{1}{\mathcal{D}_{3}}$, we proceed as follows
\begin{align}
\frac{1}{\mathcal{D}_{3}} &=-\frac{1}{r^{2}L_{30}\Big[1+\frac{1}{L_{30}} \sum_{k=1}^{4}L_{3k} x^{k}\Big]} \nonumber \\
&=-\frac{1}{r^{2}L_{30}} \Big[1+\alpha(x)\Big]^{-1}
\end{align}
where,
\begin{equation}
\alpha(x)= \frac{1}{L_{30}} \sum_{k=1}^{4}L_{3k} x^{k}
\end{equation}
For $x \ll 1$, $\alpha(x)= \frac{1}{L_{30}} \sum_{k=1}^{4}L_{3k} x^{k} \ll 1$. Taking the binomial expansion Eq. (A.52) and evaluating the terms up to the fifth order in $\alpha$ and $x$ we find
\begin{align}
\alpha^{2} &=(\frac{L_{31}^{2}}{L_{30}^{2}})x^{2}+\frac{2L_{31}L_{32}}{L_{30}^{2}}x^{3}+(\frac{2L_{31}L_{33}}{L_{30}^{2}}+\frac{L_{32}^{2}}{L_{30}^{}{2}})x^{4}+(\frac{2L_{31}L_{34}}{L_{30}^{2}}+\frac{2L_{32}L_{33}}{L_{30}^{2}})x^{5}+ ...... \\
\alpha^{3} &=\frac{L_{31}^{3}}{L_{30}^{3}}x^{3}+(\frac{3L_{31}^{2}L_{32}}{L_{30}^{3}}+\frac{3L_{31}L_{32}^{2}}{L_{30}^{3}})x^{4}+\frac{3L_{31}^{2}L_{33}}{L_{30}^{3}}x^{5}+ ....... \\
\alpha^{4} &=\frac{L_{31}^{4}}{L_{30}^{4}}x^{4}+\frac{4L_{31}^{3}L_{32}}{L_{30}^{4}}x^{5}+ ...... \\
\alpha^{5} &= \frac{L_{31}^{5}}{L_{30}^{5}} x^{5}+ .... 
\end{align}
Substituting Eqs. (A.62) to (A.66) into Eq. (A.61), we get
\begin{equation}
\frac{1}{\mathcal{D}_{3}}=-\frac{1}{r^{2}L_{30}} \Big[1+\sum_{m=1}^{5}k_{3m}x^{m}\Big],
\end{equation}
where the expansion coefficients $k_{3m}$ are listed in Table 7. \\
The first component of the eigenvector $\vec{V}_{3}$ is computed as,
\begin{align}
V_{31} &=\frac{\Delta p (\gamma+r)}{\mathcal{D}_{3}} \nonumber \\
&=-\frac{p}{L_{30}}\frac{\Delta r(1+x)}{r^{2}}\Big[1+\sum_{m=1}^{5}k_{3m}x^{m}\Big] \nonumber \\
&=-\frac{p}{L_{30}}\Big(\frac{\Delta}{\gamma}\Big)\frac{1}{\bar{n}} \sum_{m=0}^{5}a_{3m}x^{m}
\end{align} 
where the coefficients $a_{3m}$ are listed in Table 7. Proceeding in the same way, the second component of the eigenvector $\vec{V}_{3}$ is evaluated as
\begin{equation}
V_{32}=\frac{1}{L_{30}}\Big(\frac{\Delta}{\gamma}\Big) \frac{1}{\bar{n}} \sum_{m=0}^{8}b_{3m}x^{m}
\end{equation}
where the expansion coefficients $b_{3m}$ are listed in Table 7. The third component of the eigenvector $\vec{V}_{3}$ is $V_{33}=1$. 
The third eigenvector is thus
\begin{equation}
\vec{V}_{3} = 
\begin{bmatrix}
     -\frac{p}{L_{30}}\Big(\frac{\Delta}{\gamma}\Big) \frac{1}{\bar{n}} \sum_{m=0}^{5}a_{3m}x^{m}\\
    \frac{1}{L_{30}}\Big(\frac{\Delta}{\gamma}\Big) \frac{1}{\bar{n}} \sum_{m=0}^{5}b_{3m}x^{m}  \\
       1 \\
\end{bmatrix}
\end{equation}
Combining the expressions for the eigenvectors $\vec{V_{1}}$, $\vec{V_{2}}$ and $\vec{V_{3}}$ we obtain the matrix of eigenvectors of $\mathbf{A}$ as \\
\begin{equation}
\bf{M} = 
\begin{bmatrix}
    -\frac{p}{L_{12}}\Big(\frac{\Delta}{\gamma}\Big)\bar{n} \sum_{m=0}^{7}a_{1m}x^{m} & -\frac{p}{L_{22}}\Big(\frac{\Delta}{\gamma}\Big)\bar{n} \sum_{m=0}^{7}a_{2m}x^{m} & -\frac{p}{L_{30}}\Big(\frac{\Delta}{\gamma}\Big) \frac{1}{\bar{n}} \sum_{m=0}^{5}a_{3m}x^{m} \\
    \frac{1}{L_{12}} \Big(\frac{\Delta}{\gamma}\Big) \bar{n}\sum_{m=0}^{8}b_{1m}x^{m}
 & \frac{1}{L_{22}} \Big(\frac{\Delta}{\gamma}\Big) \bar{n}\sum_{m=0}^{8}b_{2m}x^{m}  & \frac{1}{L_{30}}\Big(\frac{\Delta}{\gamma}\Big) \frac{1}{\bar{n}} \sum_{m=0}^{5}b_{3m}x^{m}  \\
       1 & 1 & 1 \\
\end{bmatrix}
\end{equation} 

\subsubsection{The determinant and the inverse of the eigenvector matrix M}
Expanding Eq. (A.71) through second order in $x \ll 1$ and neglecting the insignificant terms, we get
\begin{equation}
\bf{M} = 
\begin{bmatrix}
    -\frac{p}{L_{12}}(\frac{\Delta}{\gamma})\bar{n} (a_{10}+a_{12}x^{2}) & -\frac{p}{L_{22}}(\frac{\Delta}{\gamma})\bar{n} a_{20} & -\frac{p}{L_{30}}(\frac{\Delta}{\gamma})(\frac{1}{\bar{n}}) a_{30} \\
    \frac{1}{L_{12}}(\frac{\Delta}{\gamma})\bar{n} (b_{10}+b_{12}x^{2})
 & \frac{1}{L_{22}}(\frac{\Delta}{\gamma})\bar{n} b_{20}  & \frac{1}{L_{30}}(\frac{\Delta}{\gamma})(\frac{1}{\bar{n}}) b_{30}  \\
       1 & 1 & 1 \\
\end{bmatrix}
\end{equation}
We take,
\begin{equation}
\bf{M} = 
\begin{bmatrix}
    V_{11} & V_{21} & V_{31}  \\
    V_{12} & V_{22} & V_{32}  \\
    V_{13} & V_{23} & V_{33}  \\ 
\end{bmatrix}  
\end{equation}
where the components $V_{ij}$; $i, j = 1, 2, 3$ are listed in Table 9. \\
To simplify the expression for $V_{11}$ in Eq. (A.72), we begin with the $L_{12}$ term in the denominator
\begin{align}
L_{12} &= z_{11}^{2}+2z_{10}z_{12}+2z_{11}+4z_{12}+(1-p^{2}) \nonumber \\
&= F_{11}(p)\Big(\frac{\Delta}{\gamma}\Big)^{2}+F_{12}(p)
\end{align}
where the terms $F_{11}(p)$, $F_{12}(p)$ are listed in Table 8. \\
We find that $F_{11}(p)>>F_{12}(p)$ for all $p$ and hence \\
\begin{equation}
L_{12} \approx F_{11}\bigg(\frac{\Delta}{\gamma}\bigg)^{2}
\end{equation} 
To further simplify Eq. (A.72) we substitute the expressions for $a_{10}$ and $a_{12}$ from Table 6 to the expression for $V_{11}$ in Eq. (A.72) and use Eq. (A.75) to get
\begin{align}
a_{10} &= 1 \\ 
a_{12} &=k_{11}+k_{12} \nonumber \\
&=-\frac{L_{13}}{L_{12}}-\frac{L_{14}}{L_{12}}+\bigg(\frac{L_{13}}{L_{12}}\bigg)^{2} \nonumber \\
&\approx -\frac{L_{14}}{L_{12}} \nonumber \\
&= -\frac{f_{11}}{(4+2z_{10})}\bigg(\frac{\Delta}{\gamma}\bigg)^{2}
\end{align} 
Using
\begin{equation}
\frac{a_{10}+a_{12}x^{2}}{L_{2}} = \frac{1-\frac{f_{11}}{(4+2z_{10})}\Big(\frac{\Delta}{\gamma}\Big)^{2} \frac{1}{\bar{n}^{2}}}{F_{11}(p) \Big(\frac{\Delta}{\gamma}\Big)^{2}} \\
\end{equation}
we finally obtain a simplified expression for $V_{11}$ 
\begin{align}
V_{11} &= -p \bar{n} \bigg(\frac{\Delta}{\gamma}\bigg) \Bigg[\frac{1-\frac{f_{11}}{(4+2z_{10})}(\frac{\Delta}{\gamma})^{2}(\frac{1}{\bar{n}})^{2}}{F_{11}(p)(\frac{\Delta}{\gamma})^{2}}\Bigg] \nonumber \\
&= p\bigg[-\frac{1}{F_{11}}+\frac{f_{11}}{F_{11}}\frac{1}{(2z_{10}+4)}\frac{1}{\bar{n}^{2}}\bigg(\frac{\Delta}{\gamma}\bigg)^{2}\bigg]\bar{n}\bigg(\frac{\Delta}{\gamma}\bigg)^{-1} 
\end{align}
Proceeding in a similar way we obtain analytic expressions for the other matrix elements $V_{ij}$ listed in Table~9. \\
Now that we found the analytic expression for the elements of matrix $\bf{M}$, we need to find its inverse. 
\begin{equation}
\mathbf{M}^{-1} = \frac{1}{det(\mathbf{M})} adj(\mathbf{M})
\end{equation}
where $det(\bf{M})$ is the determinant of $\bf{M}$ and $adj(\bf{M})$ is the adjoint of $\bf{M}$.\\
Using the expressions of $V_{ij}$ in Table 9, we evaluate the minors of $\bf{M}$  $T_{ij}$ in order to find its adjoint
\begin{align}
T_{11} &= V_{22}V_{33}-V_{23}V_{32} \nonumber \\
&=\frac{(3+z_{20})}{F21}\bar{n}\Big(\frac{\Delta}{\gamma}\Big)^{-1}-\frac{(3+z_{30})}{L_{30}}\frac{1}{\bar{n}}\Big(\frac{\Delta}{\gamma}\Big) \nonumber \\
&=\bigg[\frac{(3+z_{20})}{F21}-\frac{(3+z_{30})}{F31}\frac{1}{\bar{n}^{2}}\bigg(\frac{\Delta}{\gamma}\bigg)^{2}\bigg]\bar{n}\bigg(\frac{\Delta}{\gamma}\bigg)^{-1} \nonumber \\
&=\bigg[m_{1}(p)+m_{2}(p)\frac{1}{\bar{n}^{2}}\bigg(\frac{\Delta}{\gamma}\bigg)^{2}\bigg]\bar{n}\bigg(\frac{\Delta}{\gamma}\bigg)^{-1} 
\end{align}
where the coefficients $m_{1}(p)$, $m_{2}(p)$ are listed in Table 11.\\
Proceeding in the same way, we evaluate the remaining minors of $\bf{M}$ i.e. $T_{ij}$ which are listed in Table 10. The coefficients $m_{i}(p)$ that define $T_{ij}$ are listed in Table 11. All of the coefficients $m_{i}(p)$, ($i$ = 1 to 16) are functions of $p$ only. The adjoint of $\bf{M}$ is given by 
\begin{align}
adj(\mathbf{M}) &= \mathbf{M}^{\rm T} \nonumber \\ 
&= 
\begin{bmatrix}
    T_{11} & T_{12} & T_{13} \\
    T_{21} & T_{22} & T_{23} \\
    T_{31} & T_{32} & T_{33} \\
\end{bmatrix},
\end{align} 
and the determinant of $\bf{M}$ is
\begin{equation}
det(\mathbf{M}) = V_{11}T_{11}+V_{21}T_{21}+V_{31}T_{31}
\end{equation}

\subsection{Density Matrix Evolution: Analytical expressions in the overdamped regime [$\Delta / (\bar{n} \gamma) < f(p)$]}
In this section, we derive analytic expressions for the time evolution of the density matrix. The general solution of the Bloch-Redfield equations (Eq. (3) of the main text) can be obtained from Duhamel's formula [Ref. 23 of the main text]
\begin{equation}
\vec{\mathbf{x}}(t)=e^{\mathbf{A} t} \vec{\mathbf{x}}_{0}+\int_{0}^{t} e^{\mathbf{A} (t-s)} \vec{\mathbf{d}} \enspace ds
\end{equation}
where,
\begin{equation}
\vec{\mathbf{x}}(t) = \begin{bmatrix}
         \rho_{aa} &   \rho_{ab}^{R} & \rho_{ab}^{I}  
        \end{bmatrix}^{T}
\end{equation}
\begin{equation}
\vec{\mathbf{x}}_{0}=\begin{bmatrix}
         0 &  0 & 0  
        \end{bmatrix}^{T} 
\end{equation} is the initial vector, and 
\begin{equation}
\vec{\mathbf{d}}=\begin{bmatrix}
         r &  pr  & 0
        \end{bmatrix}^{T}
\end{equation}
is the driving vector. To solve Eq. (A.84) we need the exponential of the coefficient matrix $\bf{A}$
\begin{align}
e^{t\bf{A}} &= \mathbf{M} e^{t \mathbf{\Lambda}} \mathbf{M}^{-1} \nonumber \\
&=\frac{1}{det(\bf{M})}
\begin{bmatrix}
    V_{11} & V_{21} & V_{31} \\
    V_{12} & V_{22} & V_{32} \\
    V_{13} & V_{23} & V_{33} \\
\end{bmatrix}
\begin{bmatrix}
    e^{\lambda_{1}t} & 0 & 0 \\
    0 & e^{\lambda_{2}t} & 0 \\
    0 & 0 & e^{\lambda_{3}t} \\
\end{bmatrix}
\begin{bmatrix}
    T_{11} & T_{12} & T_{13} \\
    T_{21} & T_{22} & T_{23} \\
    T_{31} & T_{32} & T_{33} \\
\end{bmatrix} \nonumber \\ 
&=\frac{1}{det(\bf{M})}
\begin{bmatrix}
    \phi_{11} & \phi_{21} & \phi_{31} \\
    \phi_{12} & \phi_{22} & \phi_{32} \\
    \phi_{13} & \phi_{23} & \phi_{33} \\
\end{bmatrix}
\end{align}
where $\bf{M}$ is the eigenvector matrix found in the previous section, $\bf{\Lambda} = \begin{bmatrix}
    \lambda_{1} & 0 & 0 \\
    0 & \lambda_{2} & 0 \\
    0 & 0 & \lambda_{3} \\
\end{bmatrix}$ is the eigenvalue matrix and 
\begin{equation}
\phi_{ij}=\sum_{k=1}^{3}e^{\lambda_{k}t}V_{kj}T_{ki} \enspace ; \enspace i,\enspace j =\enspace 1-3.
\end{equation}
Using Eqs. (A.85) to (A.88) into Eq. (A.84), we get
\begin{align}
\vec{\mathbf{x}}(t) &= \frac{1}{det(\bf{M})} \int_{0}^{t} 
\begin{bmatrix}
\phi_{11} & \phi_{21} & \phi_{31} \\
\phi_{12} & \phi_{22} & \phi_{32} \\
\phi_{13} & \phi_{23} & \phi_{33} \\
\end{bmatrix}
\begin{bmatrix}
    r \\
    pr \\
    0 \\
\end{bmatrix}  
ds  \nonumber \\  
&=\frac{r}{det(\bf{M})} \int_{0}^{t}
\begin{bmatrix}
\phi_{11} + p \phi_{21} \\
\phi_{12} + p \phi_{22} \\
\phi_{13} + p \phi_{23} \\
\end{bmatrix} 
ds  
\end{align} 
Evaluating the integrals of $\vec{\mathbf{x}}(t)$ from Eq. (A.90), we find the expressions for $\rho_{aa}(t)$,$\rho_{ab}^{R}(t)$, and $\rho_{ab}^{I}(t)$
\begin{align}
\rho_{aa}(t) &=\frac{r}{det(\bf{M})} \sum_{k=1}^{3}\frac{(1-e^{\lambda_{k}t})}{-\lambda_{k}}V_{k1}(T_{k1}+pT_{k2}) \\
\rho_{ab}^{R}(t) &=\frac{r}{det(\bf{M})} \sum_{k=1}^{3}\frac{(1-e^{\lambda_{k}t})}{-\lambda_{k}}V_{k2}(T_{k1}+pT_{k2}) \\
\rho_{ab}^{I}(t) &=\frac{r}{det(\bf{M})} \sum_{k=1}^{3}\frac{(1-e^{\lambda_{k}t})}{-\lambda_{k}}V_{k3}(T_{k1}+pT_{k2})
\end{align} 
To express these general solutions in terms of the physical parameters, we use Eq. (A.83) for the determinant of matrix $\bf{M}$ and the  coefficients $V_{ij}$, $T_{ij}$ listed in Tables 9 and 10. The determinant of matrix $\bf{M}$ is obtained from Eq. (A.83) using 
\begin{equation}
\begin{split}
det(\mathbf{M}) = p\Big[\Big(-\frac{m_{1}}{F_{11}}+\Big(-\frac{m_{2}}{F_{11}}+\frac{f_{11}}{F_{11}}\frac{1}{(2z_{10}+4)}\Big)\frac{1}{\bar{n}^{2}}\Big(\frac{\Delta}{\gamma}\Big)^{2}\Big)+p\Big(-\frac{m_{5}}{F_{21}}-\frac{m_{6}}{F_{21}}\frac{1}{\bar{n}^{2}}\Big(\frac{\Delta}{\gamma}\Big)^{2}\Big)\\
+p\Big(-\frac{m_{11}}{L_{30}}\frac{1}{\bar{n}^{2}}\Big(\frac{\Delta}{\gamma}\Big)^{2}-\frac{m_{12}}{L_{30}}\frac{1}{\bar{n}^{4}}\Big(\frac{\Delta}{\gamma}\Big)^{4}\Big)\Big]\bar{n}^{2}\Big(\frac{\Delta}{\gamma}\Big)^{-2} \\
\end{split}
\end{equation}
The term proportional to $\frac{1}{\bar{n}^{4}}(\frac{\Delta}{\gamma})^{4})$ can be neglected for large $\bar{n}$ and we find
\begin{align}
det(\bf{M}) &= p\Big[\Big(-\frac{m_{1}}{F_{11}}-\frac{m_{5}}{F_{21}}\Big)+\Big(-\frac{m_{2}}{F_{11}}+\frac{f_{11}}{F_{11}}\frac{1}{(2z_{10}+4)}-\frac{m_{6}}{F_{21}}-\frac{m_{11}}{L_{30}})\frac{1}{\bar{n}^{2}}\Big(\frac{\Delta}{\gamma}\Big)^{2}\Big]\bar{n}^{2}\Big(\frac{\Delta}{\gamma}\Big)^{-2} \nonumber \\
&=\Big[T_{1}(p)+T_{2}(p)\frac{1}{\bar{n}^{2}}\Big(\frac{\Delta}{\gamma}\Big)^{2}\Big]\bar{n}^{2}\Big(\frac{\Delta}{\gamma}\Big)^{-2}
\end{align}
where the coefficients $T_{1}(p)$, $T_{2}(p)$ are listed in Table 8. \\
To find the expression for $\rho_{aa}(t)$, we calculate the following terms in Eq. (A.91),      
\begin{equation}
V_{11}(T_{11}+pT_{12}) = \Big[A_{1}+A_{2}\frac{1}{\bar{n}^{2}}\Big(\frac{\Delta}{\gamma}\Big)^{2}\Big]\bar{n}^{2}\Big(\frac{\Delta}{\gamma}\Big)^{-2}
\end{equation}
\begin{equation}
V_{21}(T_{21}+pT_{22}) = \Big[A_{3}+A_{4}\frac{1}{\bar{n}^{2}}\Big(\frac{\Delta}{\gamma}\Big)^{2}\Big]\bar{n}^{2}\Big(\frac{\Delta}{\gamma}\Big)^{-2}
\end{equation} 
\begin{equation}
V_{31}(T_{31}+pT_{32}) = \Big[A_{5}+A_{6}\frac{1}{\bar{n}^{2}}\Big(\frac{\Delta}{\gamma}\Big)^{2}\Big] 
\end{equation}
where the coefficients $A_{i}(p)$ [$i = 1 - 6$] are listed in Table 12. All the terms $A_{i}(p)$ depend on $p$ only.
Using Eqs. (96) to (98) in Eq. (91) the general expression of $\rho_{aa}(t)$ can be recast in the form\\
\begin{equation}
\begin{split}
\rho_{aa}(t)= \frac{r}{det(\bf{M})}\Big[\Big(A_{1}+A_{2}\frac{1}{\bar{n}^{2}}\Big(\frac{\Delta}{\gamma}\Big)^{2}\Big)\bar{n}^{2}\Big(\frac{\Delta}{\gamma}\Big)^{-2}\Big(\frac{1-e^{\lambda_{1}t}}{-\lambda_{1}}\Big)+\Big(A_{3}+A_{4}\frac{1}{\bar{n}^{2}}\Big(\frac{\Delta}{\gamma}\Big)^{2}\Big)\bar{n}^{2}\Big(\frac{\Delta}{\gamma}\Big)^{-2}\Big(\frac{1-e^{\lambda_{2}t}}{-\lambda_{2}}\Big)\\+
\Big(A_{5}+A_{6}\frac{1}{\bar{n}^{2}}\Big(\frac{\Delta}{\gamma}\Big)^{2}\Big)\Big(\frac{1-e^{\lambda_{3}t}}{-\lambda_{3}}\Big)\Big]
\end{split}
\end{equation}
Proceeding in the same way as for $\rho_{aa}(t)$, we find the expressions for $\rho^{R}_{ab}(t)$ and $\rho^{I}_{ab}(t)$ as        
\begin{equation}
\begin{split}
\rho^{R}_{ab}(t)= \frac{r}{det(\bf{M})}\Big[\Big(B_{1}+B_{2}\frac{1}{\bar{n}^{2}}\Big(\frac{\Delta}{\gamma}\Big)^{2}\Big)\bar{n}^{2}\Big(\frac{\Delta}{\gamma}\Big)^{-2}\Big(\frac{1-e^{\lambda_{1}t}}{-\lambda_{1}}\Big)+\Big(B_{3}+B_{4}\frac{1}{\bar{n}^{2}}\Big(\frac{\Delta}{\gamma}\Big)^{2}\Big)\bar{n}^{2}\Big(\frac{\Delta}{\gamma}\Big)^{-2}\Big(\frac{1-e^{\lambda_{2}t}}{-\lambda_{2}}\Big)\\+
\Big(B_{5}+B_{6}\frac{1}{\bar{n}^{2}}\Big(\frac{\Delta}{\gamma}\Big)^{2}\Big)\Big(\frac{1-e^{\lambda_{3}t}}{-\lambda_{3}}\Big)\Big]
\end{split}
\end{equation}
\begin{equation}
\begin{split}
\rho^{I}_{ab}(t) = \frac{r}{det(\bf{M})}\Big[\Big(C_{1}+C_{2}\frac{1}{\bar{n}^{2}}\Big(\frac{\Delta}{\gamma}\Big)^{2}\Big)\Big(\frac{1-e^{\lambda_{1}t}}{-\lambda_{1}}\Big)+\Big(C_{3}+C_{4}\frac{1}{\bar{n}^{2}}\Big(\frac{\Delta}{\gamma}\Big)^{2}\Big)\Big(\frac{1-e^{\lambda_{2}t}}{-\lambda_{2}}\Big) \\ +\Big
(C_{5}+C_{6}\frac{1}{\bar{n}^{2}}\Big(\frac{\Delta}{\gamma}\Big)^{2}\Big)\Big(\frac{1-e^{\lambda_{3}t}}{-\lambda_{3}}\Big)\Big]\bar{n}\Big(\frac{\Delta}{\gamma}\Big)^{-1}
\end{split}
\end{equation}
where the coefficients $B_{i}(p)$, $C_{i}(p)$, [$i = 1 - 6$] are listed in Table 12. All the terms $B_{i}(p)$ and $C_{i}(p)$ depend on $p$ only.\\
The general expressions for $\rho_{aa}(t)$, $\rho^{R}_{ab}(t)$ and $\rho^{I}_{ab}(t)$ can take two explicit forms depending on the value of the alignment factor $p$. If the alignment factor is greater than the critical value $i.e.$ $p>p_{c}$, then
\begin{align}
\lambda_{j} &= \gamma z_{j0} \bar{n} \nonumber \\
&= -r |z_{j0}| \enspace ;\enspace j =\enspace 1,\enspace 3
\end{align}
However the second eigenvalue has a different scaling relation as shown in Sec II.C of the main text
\begin{equation}
\lambda_{2} = -\gamma |f_{21}| \frac{1}{\bar{n}}\bigg(\frac{\Delta}{\gamma}\bigg)^{2} 
\end{equation}
Substituting the eigenvalues from Eqs. (A.102) and (A.103) into Eqs. (A.99) to (A.101), using Eq. (A.95) for det($\mathbf{M}$) and neglecting the small term proportional to $\frac{1}{\bar{n}^{4}}(\frac{\Delta}{\gamma})^{4}$ for large $\bar{n}$, we get
\begin{equation}
\begin{split}	
\rho_{aa}(t) = \frac{1}{\Big[T_{1}(p)+T_{2}(p)\frac{1}{\bar{n}^{2}}\Big(\frac{\Delta}{\gamma}\Big)^{2}\Big]}\Bigg[\Big(A_{1}+A_{2}\frac{1}{\bar{n}^{2}}\Big(\frac{\Delta}{\gamma}\Big)^{2}\Big)\Big(\frac{1-e^{-\gamma|z_{10}|\bar{n}t}}{ |z_{10}|}\Big)
+\Big(A_{3}+A_{4}\frac{1}{\bar{n}^{2}}\Big(\frac{\Delta}{\gamma}\Big)^{2}\Big)\bar{n}^{2}\Big(\frac{\Delta}{\gamma}\Big)^{-2} \\ \times \Big(\frac{1-e^{-\gamma|f_{21}|\frac{1}{\bar{n}}\Big(\frac{\Delta}{\gamma}\Big)^{2}t}}{|f_{21}|}\Big)+
A_{5}\frac{1}{\bar{n}^{2}}\Big(\frac{\Delta}{\gamma}\Big)^{2}\Big(\frac{1-e^{-\gamma|z_{30}|\bar{n}t}}{ |z_{30}|}\Big)\Bigg] 
\end{split}
\end{equation}
\begin{equation}
\begin{split}
\rho^{R}_{ab}(t)= \frac{1}{\Big[T_{1}(p)+T_{2}(p)\frac{1}{\bar{n}^{2}}\Big(\frac{\Delta}{\gamma}\Big)^{2}\Big]}\Bigg[\Big(B_{1}+B_{2}\frac{1}{\bar{n}^{2}}\Big(\frac{\Delta}{\gamma}\Big)^{2}\Big)\Big(\frac{1-e^{-\gamma|z_{10}|\bar{n}t}}{ |z_{10}|}\Big)
+\Big(B_{3}+B_{4}\frac{1}{\bar{n}^{2}}\Big(\frac{\Delta}{\gamma}\Big)^{2}\Big)\bar{n}^{2}\Big(\frac{\Delta}{\gamma}\Big)^{-2} \\ \times \Big(\frac{1-e^{-\gamma|f_{21}|\frac{1}{\bar{n}}\Big(\frac{\Delta}{\gamma}\Big)^{2}t}}{|f_{21}|}\Big)+
B_{5}\frac{1}{\bar{n}^{2}}\Big(\frac{\Delta}{\gamma}\Big)^{2}\Big(\frac{1-e^{-\gamma|z_{30}|\bar{n}t}}{ |z_{30}|}\Big)\Bigg] 
\end{split}
\end{equation}
\begin{equation}
\begin{split}
\rho^{I}_{ab}(t)= \frac{1}{\Big[T_{1}(p)+T_{2}(p)\frac{1}{\bar{n}^{2}}\Big(\frac{\Delta}{\gamma}\Big)^{2}\Big]} \frac{1}{\bar{n}} \Big(\frac{\Delta}{\gamma}\Big)\Bigg[\Big(C_{1}+C_{2}\frac{1}{\bar{n}^{2}}\Big(\frac{\Delta}{\gamma}\Big)^{2}\Big)\Big(\frac{1-e^{-\gamma|z_{10}|\bar{n}t}}{ |z_{10}|}\Big)
+\Big(C_{3}+C_{4}\frac{1}{\bar{n}^{2}}\Big(\frac{\Delta}{\gamma}\Big)^{2}\Big)\bar{n}^{2}\Big(\frac{\Delta}{\gamma}\Big)^{-2} \\ \times \Big(\frac{1-e^{-\gamma|f_{21}|\frac{1}{\bar{n}}\Big(\frac{\Delta}{\gamma}\Big)^{2}t}}{|f_{21}|}\Big)+
\Big(C_{5}+C_{6}\frac{1}{\bar{n}^{2}}\Big(\frac{\Delta}{\gamma}\Big)^{2}\Big)\Big(\frac{1-e^{-\gamma|z_{30}|\bar{n}t}}{ |z_{30}|}\Big)\Bigg]
\end{split}
\end{equation}
Equations (A.104), (A.105), and (A.106) are identical to Eqs. (18), (19) and (20) of the main text. \\
If the alignment factor is less than the critical value ($p<p_{c}$), then
\begin{align}
\lambda_{j} &= \gamma z_{j0} \bar{n} \nonumber \\
&= -r |z_{j0}| \enspace ;\enspace j =\enspace 1,\enspace 2,\enspace 3
\end{align}
Substituting the eigenvalues from Eq. (A.107) into Eqs. (A.99) to (A.101), using Eq. (A.95) for det($\mathbf{M}$) and neglecting the small term proportional to $\frac{1}{\bar{n}^{4}}(\frac{\Delta}{\gamma})^{4}$ for large $\bar{n}$, we get
\begin{equation}
\begin{split}
\rho_{aa}(t) = \frac{1}{\Big[T_{1}(p)+T_{2}(p)\frac{1}{\bar{n}^{2}}\Big(\frac{\Delta}{\gamma}\Big)^{2}\Big]}\Bigg[\Big(A_{1}+A_{2}\frac{1}{\bar{n}^{2}}\Big(\frac{\Delta}{\gamma}\Big)^{2}\Big)\Big(\frac{1-e^{-\gamma|z_{10}|\bar{n}t}}{ |z_{10}|}\Big)
+\Big(A_{3}+A_{4}\frac{1}{\bar{n}^{2}}\Big(\frac{\Delta}{\gamma}\Big)^{2}\Big)\Big(\frac{1-e^{-\gamma|z_{20}|\bar{n}t}}{ |z_{20}|}\Big)+ \\ 
A_{5}\frac{1}{\bar{n}^{2}}\Big(\frac{\Delta}{\gamma}\Big)^{2}\Big(\frac{1-e^{-\gamma|z_{30}|\bar{n}t}}{ |z_{30}|}\Big)\Bigg] 
\end{split}
\end{equation}
\begin{equation}
\begin{split}
\rho^{R}_{ab}(t)= \frac{1}{\Big[T_{1}(p)+T_{2}(p)\frac{1}{\bar{n}^{2}}\Big(\frac{\Delta}{\gamma}\Big)^{2}\Big]}\Bigg[\Big(B_{1}+B_{2}\frac{1}{\bar{n}^{2}}\Big(\frac{\Delta}{\gamma}\Big)^{2}\Big)\Big(\frac{1-e^{-\gamma|z_{10}|\bar{n}t}}{ |z_{10}|}\Big)
+\Big(B_{3}+B_{4}\frac{1}{\bar{n}^{2}}\Big(\frac{\Delta}{\gamma}\Big)^{2}\Big)\Big(\frac{1-e^{-\gamma|z_{20}|\bar{n}t}}{ |z_{20}|}\Big)+ \\
B_{5}\frac{1}{\bar{n}^{2}}\Big(\frac{\Delta}{\gamma}\Big)^{2}\Big(\frac{1-e^{-\gamma|z_{30}|\bar{n}t}}{ |z_{30}|}\Big)\Bigg]
\end{split}
\end{equation} \\
\begin{equation}
\begin{split}
\rho^{I}_{ab}(t)= \frac{1}{\Big[T_{1}(p)+T_{2}(p)\frac{1}{\bar{n}^{2}}\Big(\frac{\Delta}{\gamma}\Big)^{2}\Big]} \frac{1}{\bar{n}} \Big(\frac{\Delta}{\gamma}\Big)\Bigg[(C_{1}+C_{2}\frac{1}{\bar{n}^{2}}\Big(\frac{\Delta}{\gamma}\Big)^{2}\Big)\Big(\frac{1-e^{-\gamma|z_{10}|\bar{n}t}}{ |z_{10}|}\Big)+\Big(C_{3}+C_{4}\frac{1}{\bar{n}^{2}}\Big(\frac{\Delta}{\gamma}\Big)^{2}\Big)\Big(\frac{1-e^{-\gamma|z_{20}|\bar{n}t}}{ |z_{20}|}\Big)\\+\Big
(C_{5}+C_{6}\frac{1}{\bar{n}^{2}}\Big(\frac{\Delta}{\gamma}\Big)^{2}\Big)\Big(\frac{1-e^{-\gamma|z_{30}|\bar{n}t}}{ |z_{30}|}\Big)\Bigg] 
\end{split}
\end{equation}
Equations (A.108), (A.109) and (A.110) are identical to Eqs. (24), (25) and (26) of the main text. \\
In the limit of small energy level spacing ($\frac{\Delta}{\gamma} \ll 1$) and large $\bar{n}$, we can neglect the terms proportional to $\frac{1}{\bar{n}^{2}}(\frac{\Delta}{\gamma})^{-2}$ in the above equations and the general solution further simplifies. For $p>p_{c}$, we obtain
\begin{align}
\rho_{aa}(t) &= \frac{1}{T_{1}(p)}\Bigg[A_{1}\Big(\frac{1-e^{-\gamma|z_{10}|\bar{n}t}}{ |z_{10}|}\Big)
+A_{4}\Big(\frac{1-e^{-\gamma|f_{21}|\frac{1}{\bar{n}}\big(\frac{\Delta}{\gamma}\big)^{2}t}}{|f_{21}|}\Big)\Bigg] \\
\rho^{R}_{ab}(t) &= \frac{1}{T_{1}(p)}\Bigg[B_{1}\Big(\frac{1-e^{-\gamma|z_{10}|\bar{n}t}}{ |z_{10}|}\Big)
+B_{4}\Big(\frac{1-e^{-\gamma|f_{21}|\frac{1}{\bar{n}}\big(\frac{\Delta}{\gamma}\big)^{2}t}}{|f_{21}|}\Big)\Bigg]
\end{align}
\begin{equation}
\begin{split}
\rho^{I}_{ab}(t) = \frac{1}{T_{1}(p)}  \Big(\frac{\Delta}{\bar{n}\gamma}\Big)\Bigg[C_{1}\Big(\frac{1-e^{-\gamma|z_{10}|\bar{n}t}}{ |z_{10}|}\Big)
+C_{4}\Big(\frac{1-e^{-\gamma|f_{21}|\frac{1}{\bar{n}}\big(\frac{\Delta}{\gamma}\big)^{2}t}}{|f_{21}|}\Big) +C_{5}\Big(\frac{1-e^{-\gamma|z_{30}|\bar{n}t}}{ |z_{30}|}\Big)\Bigg]
\end{split}
\end{equation}
Equations (A.111), (A.112) and (A.113) are the same as Eqs. (27), (28) and (29) of the main text. \\
For $p<p_{c}$, we find
\begin{align}
\rho_{aa}(t) &= \frac{1}{T_{1}(p)}\Bigg[A_{1}\Big(\frac{1-e^{-\gamma|z_{10}|\bar{n}t}}{ |z_{10}|}\Big)+A_{3} \Big(\frac{1-e^{-\gamma|z_{20}|\bar{n}t}}{ |z_{20}|}\Big)\Bigg] \\
\rho^{R}_{ab}(t) &= \frac{1}{T_{1}(p)}\Bigg[B_{1}\Big(\frac{1-e^{-\gamma|z_{10}|\bar{n}t}}{ |z_{10}|}\Big)
+B_{3}\Big(\frac{1-e^{-\gamma|z_{20}|\bar{n}t}}{ |z_{20}|}\Big)\Bigg] \\
\rho^{I}_{ab}(t) &= \frac{1}{T_{1}(p)} \Big(\frac{\Delta}{\bar{n}\gamma}\Big)\Bigg[C_{1}\Big(\frac{1-e^{-\gamma|z_{10}|\bar{n}t}}{ |z_{10}|}\Big)+C_{3}\Big(\frac{1-e^{-\gamma|z_{20}|\bar{n}t}}{ |z_{20}|}\Big)+C_{5}\Big(\frac{1-e^{-\gamma|z_{30}|\bar{n}t}}{ |z_{30}|}\Big)\Bigg]
\end{align}
The above Eqs. (A.114), (A.115) and (A.116) are the same as Eqs. (33), (34) and (35) of the main text.

}

{\fontfamily{cmr}\selectfont \tiny

\begin{table}
\begin{tabular}{ |p{2.5cm}||p{8.5cm}| |p{4cm}|}
 \hline
 \multicolumn{3}{|c|}{Table 1: The expansion coefficients $c_{k}$, $d_{k}$ and $b_{k}$ for $\mathcal{D}$, $E$ and $\beta(x)$} \\
 \hline
$c_{i}$& $d_{i}$ & $b_{i}$\\
 \hline
0 & $d_{0}=4 (\frac{\Delta^{2}}{\gamma^{2}}-p^{2})^{3}$ &$K = \sqrt[3]{\frac{c_{3}}{6}+\sqrt{\frac{d_{6}}{108}}}$\\

$c_{1}=4(\frac{\Delta}{\gamma})^{2}+2p^{2}$ &$d_{1}=-48p^{2} (\frac{\Delta^{2}}{\gamma^{2}}-p^{2})^{2}$&$b_{1}=\frac{\frac{c_{2}}{6}+\sqrt{\frac{d_{6}}{108}}u_{1}}{\sqrt[3]{\frac{c_{3}}{6}+\sqrt{\frac{d_{6}}{108}}}}$\\

$c_{2}=8p^{2}$&$d_{2}=12(2\frac{\Delta^{2}}{\gamma^{2}}+p^{2})^{2}+192p^{4}(\frac{\Delta^{2}}{\gamma^{2}}-
p^{2})-4(4+9p^{2})(\frac{\Delta^{2}}{\gamma^{2}}-p^{2})^{2}$ & $b_{2}=\frac{\frac{c_{1}}{6}+\sqrt{\frac{d_{6}}{108}}u_{2}}{\sqrt[3]{\frac{c_{3}}{6}+\sqrt{\frac{d_{6}}{108}}}}$\\

$c_{3}=\frac{(16+5p^{2})}{9}$ &$d_{3}=96p^{2}(2\frac{\Delta^{2}}{\gamma^{2}}+p^{2})+32p^{2}(4+9p^{2})(\frac{\Delta^{2}}{\gamma^{2}}-p^{2})-256p^{6}$& $b_{3}=\frac{\sqrt{\frac{d_{6}}{108}}u_{3}}{\sqrt[3]{\frac{c_{3}}{6}+\sqrt{\frac{d_{6}}{108}}}}$\\

0& $d_{4}=\frac{8}{3}(8+27p^{2})(2\frac{\Delta^{2}}{\gamma^{2}}+p^{2})+\frac{4}{3}(4+9p^{2})^{2}(\frac{\Delta^{2}}{\gamma^{2}}-p^{2})-64p^{4}(1+9p^2)$&$b_{4}=\frac{\sqrt{\frac{d_{6}}{108}}u_{4}}{{\sqrt[3]{\frac{c_{3}}{6}+\sqrt{\frac{d_{6}}{108}}}}}$\\

0& $d_{5}= -16p^{4}(6+27p^{2})$ &$b_{5}=\frac{\sqrt{\frac{d_{6}}{108}}u_{5}}{{\sqrt[3]{\frac{c_{3}}{6}+\sqrt{\frac{d_{6}}{108}}}}}$\\

0&$d_{6}= -36p^{4}(1+3p^{2})$ &$b_{6}=\frac{\sqrt{\frac{d_{6}}{108}}u_{6}}{{\sqrt[3]{\frac{c_{3}}{6}+\sqrt{\frac{d_{6}}{108}}}}}$\\
 \hline
\end{tabular}\\
\end{table}

\begin{table}
\begin{tabular}{ |p{10cm}||p{6cm}| }
 \hline
 \multicolumn{2}{|c|}{Table 2: The expansion coefficients $u_{k}$ and $v_{k}$ in the expressions for $\sqrt{\mathcal{D}}$ and $\Lambda(x)$} \\
 \hline
$u_{i}$ & $v_{i}$\\
 \hline
$u_{1}=\frac{1}{2} \frac{d_{5}}{d_{6}}$& $v_{1}=\frac{b_{1}}{3}$\\

$u_{2}=\frac{1}{2}\frac{d_{4}}{d_{6}}-\frac{1}{8}(\frac{d_{5}}{d_{6}})^{2}$& $v_{2}=\frac{b_{2}}{3}-\frac{b_{1}^{2}}{9}$\\

$u_{3}=\frac{1}{2}\frac{d_{3}}{d_{6}}-\frac{1}{8}2\frac{d_{4}d_{5}}{d_{6}^{2}}+\frac{3}{48}(\frac{d_{5}}{d_{6}})^{3}$& $v_{3}=\frac{b_{3}}{3}-\frac{2b_{1}b_{2}}{9}+\frac{5b_{1}^{3}}{81}$\\

$u_{4}=\frac{1}{2}\frac{d_{2}}{d_{6}}-\frac{1}{8}({2\frac{d_{3}d_{5}}{d_{6}^{2}}+(\frac{d_{4}}{d_{6}})^{2}})+\frac{3}{48}(3\frac{d_{4}d_{5}^{2}}{d_{6}^{3}})-\frac{15}{384}(\frac{d_{5}}{d_{6}})^{4}$&$v_{4}=\frac{b_{4}}{3}-\frac{(2b_{1}b_{3}+b_{2}^{2})}{9}+\frac{5(3b_{1}^{2}b_{2})}{81}$\\

$u_{5}=\frac{1}{2}\frac{d_{1}}{d_{6}}-\frac{1}{8}(2\frac{d_{2}d_{5}}{d_{6}^{2}}+2\frac{d_{3}d_{4}}{d_{6}^{2}})+\frac{3}{48}(3\frac{d_{3}d_{5}^{2}}{d_{6}^{3}}+3\frac{d_{4}^{2}d_{5}}{d_{6}^{3}})-\frac{15}{384}(4\frac{d_{4}d_{5}^{3}}{d_{6}^{4}})$&$v_{5}=\frac{b_{5}}{3}-\frac{(2b_{1}b_{4}+2b_{2}b_{3})}{9}+\frac{5(3b_{1}^{2}b_{3}+3b_{1}b_{2}^{2})}{81}$\\

$u_{6}=\frac{1}{2}\frac{d_{0}}{d_{6}}-\frac{1}{8}(2\frac{d_{1}d_{5}}{d_{6}^{2}}+2\frac{d_{2}d_{4}}{d_{6}^{2}}+(\frac{d_{3}}{d_{6}})^{3})+\frac{3}{48}(3\frac{d_{2}d_{5}^{2}}{d_{6}^{3}}+6\frac{d_{3}d_{4}d_{5}}{d_{6}^{3}}+(\frac{d_{4}}{d_{6}})^{3})-\frac{15}{384}(4\frac{d_{3}d_{5}^{3}}{d_{6}^{4}}+6\frac{d_{4}^{2}d_{5}^{2}}{d_{6}^{4}})$&$v_{6}=\frac{b_{6}}{3}-\frac{(2b_{1}b_{5}+2b_{2}b_{4}+b_{3}^{2})}{9}+\frac{5(3b_{1}^{2}b_{4}+6b_{1}b_{2}b_{3}+b_{2}^{3})}{81}$\\
 \hline
\end{tabular}
\end{table}

\begin{table}
\begin{tabular}{ |p{7cm}||p{9cm}| }
 \hline
 \multicolumn{2}{|c|}{Table 3: The expansion coefficients $\mathcal{W}_{k}$ and $z_{jk}$ in the expression of $\frac{1}{\mathcal{T}}$ in Eq. (A.40) and $\lambda_{j}$ in Eq. (A.43)} \\
 \hline
$\mathcal{W}_{i}$ & $z_{ji}$\\
 \hline
0  & $z_{j0}=-\frac{5}{3}-\frac{\alpha_{j}}{3K}(\frac{4}{3}+3p^{2})-\beta_{j} K$\\

$\mathcal{W}_{1}=-v_{1}$& $z_{j1}=-1-\frac{\alpha_{j}}{3K}(4p^{2}+(\frac{4}{3}+3p^{2})W_{1})-\beta_{j} Kv_{1}$\\

$\mathcal{W}_{2}=-v_{2}+v_{1}^{2}$&$z_{j2}=\frac{\alpha_{j}}{3K}[(\frac{\Delta^{2}}{\gamma^{2}}-p^{2})-4p^{2}W_{1}-(\frac{4}{3}+3p^{2})W_{2})] - \beta_{j} Kv_{2}$ \\

$\mathcal{W}_{3}=-v_{3}+2v_{1}v_{2}-v_{1}^{3}$& $z_{j3}=\frac{\alpha_{j}}{3K}[(\frac{\Delta^{2}}{\gamma^{2}}-p^{2})W_{1}-4p^{2}W_{2}-(\frac{4}{3}+3p^{2})W_{3})] - \beta_{j} Kv_{3}$\\

$\mathcal{W}_{4}=-v_{4}+(2v_{1}v_{3}+v_{2}^{2})-3v_{1}^{2}v_{2}+v_{1}^{4}$&$z_{j4}=\frac{\alpha_{j}}{3K}[(\frac{\Delta^{2}}{\gamma^{2}}-p^{2})W_{2}-4p^{2}W_{3}-(\frac{4}{3}+3p^{2})W_{4})] - \beta_{j} Kv_{4}$\\

$\mathcal{W}_{5}=-v_{5}+(2v_{1}v_{4}+2v_{2}v_{3})-(3v_{1}^{2}v_{3}+3v_{1}v_{2}^{2})+4v_{1}^{3}v_{2}-v_{1}^{5}$&$z_{j5}=\frac{\alpha_{j}}{3K}[(\frac{\Delta^{2}}{\gamma^{2}}-p^{2})W_{3}-4p^{2}W_{4}-(\frac{4}{3}+3p^{2})W_{5})] - \beta_{j} Kv_{5}$\\

$\mathcal{W}_{6}=-v_{6}+(2v_{1}v_{5}+2v_{2}v_{4}+v_{3}^{2})-(3v_{1}^{2}v_{4}+6v_{1}v_{2}v_{3}+v_{3}^{2})+(4v_{1}^{3}v_{3}+6v_{1}^{2}v_{2}^{2})+v_{1}^{6}$&$z_{j6}=\frac{\alpha_{j}}{3K}[(\frac{\Delta^{2}}{\gamma^{2}}-p^{2})W_{4}-4p^{2}W_{5}-(\frac{4}{3}+3p^{2})W_{6})] - \beta_{j} Kv_{6}$\\

0 &$z_{j7}=\frac{\alpha_{j}}{3K}[(\frac{\Delta^{2}}{\gamma^{2}}-p^{2})W_{5}-4p^{2}W_{6}]$\\

0 &$z_{j8}=\frac{\alpha_{j}}{3K}[(\frac{\Delta^{2}}{\gamma^{2}}-p^{2})W_{6}]$\\
 \hline
\end{tabular}
\end{table}

\begin{table}
\begin{tabular}{ |p{9.5cm}| |p{7.5cm}| }
 \hline
 \multicolumn{2}{|c|}{Table 4: Coefficients $z_{j2}$ and $v_{2}$}\\
 \hline
$z_{j2}$ & $v_{2}$\\
 \hline
$f_{j1}(p)=\frac{\alpha_{j}}{3K}+[\frac{\alpha_{j}}{3K}(\frac{4}{3}+3p^{2})-\beta_{j} K]s_{1}(p)$ & $s_{1}(p)=\frac{l_{1}(p)}{3}$\\

$f_{j2}(p)=-\frac{\alpha_{j}}{3K} p^{2}+4p^{2}(\frac{\alpha_{j}}{3K})v_{1}+[\frac{\alpha_{j}}{3K}(\frac{4}{3}+3p^{2})-\beta_{j} K]s_{2}(p)-\frac{\alpha_{j}}{3K}(\frac{4}{3}+3p^{2})v_{1}^{2}$ & $s_{2}(p)=\frac{l_{2}(p)}{3}-[\frac{1}{3K}(\frac{4p^{2}}{3}+\frac{2}{9}\frac{p^{2}(6+27p^{2})}{(1+3p^{2})}\sqrt{\frac{(1+3p^{2})}{3}}i)]^{2}$\\
 \hline
\end{tabular}\\
\end{table}

\begin{table}
\begin{tabular}{ |p{6.25cm}| |p{6.25cm}| |p{4cm}| }
 \hline
 \multicolumn{3}{|c|}{Table 5: Coefficients $d_{4}$, $u_{2}$ and $b_{2}$ in the expansion of $z_{j2}$ }\\
 \hline
$b_{2}$ &  $u_{2}$ & $d_{4}$ \\
 \hline
$l_{1}(p)=\frac{2}{3K}+\frac{p^{2}}{K}\sqrt{\frac{(1+3p^{2})}{3}}g_{1}(p)i$ & $g_{1}(p)=-\frac{h_{1}(p)}{72p^{4}(1+3p^{2})}$ & $h_{1}(p)=4(16+60p^{2}+27p^{4})$\\

$l_{2}(p)=\frac{p^{2}}{3K}+\frac{p^{2}}{K}\sqrt{\frac{(1+3p^{2})}{3}}g_{2}(p)i$ & $g_{2}(p)=-\frac{h_{2}(p)}{72p^{4}(1+3p^{2})}-\frac{2}{81}(\frac{6+27p^{2}}{1+3p^{2}})^{2}$ & $h_{2}(p)=-4p^{4}(22+171p^{2})$\\
 \hline
\end{tabular}\\
\end{table}

\begin{table}
\begin{tabular}{ |p{4cm}| |p{3.5cm}| |p{2.5cm}| |p{6cm}|}
 \hline
 \multicolumn{4}{|c|}{Table 6: Coefficients for the eigenvectors $V_{j}$, $j$ = 1, 2 in Eq. (A.44)}\\
 \hline
$L_{jk}; j = 1,2,3$& $k_{jk}; j=1,2$ & $a_{jk}; j=1,2$ & $b_{jk}; j =1,2$\\
 \hline
$L_{j0}=z_{j0}^{2}+4z_{j0}+3(1-p^{2})$ & 0 & $a_{j0}=1$ & $b_{j0}= 3 + z_{j0}$\\

$L_{j1}=2z_{j0}z_{j1}+2z_{j0}+4z_{j1}+4(1-p^{2})$ & $k_{j1}=-\frac{L_{j3}}{L_{j2}}$ &$a_{j1}= 1 + k_{j1}$ & $b_{j1}= (1+z_{j1})+(3+z_{j0})k_{j1}$\\

$L_{j2}=z_{j1}^{2}+2z_{j0}z_{j2}+2z_{j1}+4z_{j2}+(1-p^{2})$ & $k_{j2}=-\frac{L_{j4}}{L_{j2}}+(\frac{L_{j3}}{L_{j2}})^{2}$ & $a_{j2}= k_{j1} + k_{j2}$ & $b_{j2}=(1+z_{j1})k_{j1}+(3+z_{j0})k_{j2}+z_{j2}$\\

$L_{j3}=2z_{j1}z_{j2}+2z_{j2}$ & $k_{j3}=\frac{2 L_{j3} L_{j4}}{L_{j2}^{2}}-(\frac{L_{j3}}{L_{j2}})^{3}$ & $a_{j3}= k_{j2} + k_{j3}$ & $b_{j3}=(1+z_{j1})k_{j2}+(3+z_{j0})k_{j3}+z_{j2}k_{j1}$\\

$L_{j4}=z_{j2}^{2}$ & $k_{j4}=(\frac{L_{j4}}{L_{j2}})^{2}- \frac{3 L_{j3}^{2} L_{j4}}{L_{j2}^{3}}$ & $a_{j4}= k_{j3} + k_{j4}$ & $b_{j4}=(1+z_{j1})k_{j3}+(3+z_{j0})k_{j4}+z_{j2}k_{j2}$\\

0 & $k_{j5}=-\frac{3 L_{j3} L_{j4}^{2}}{L_{j2}^{3}}$ & $a_{j5}= k_{j4} + k_{j5}$ & $b_{j5}=(1+z_{j1})k_{j4}+(3+z_{j0})k_{j5}+z_{j2}k_{j3}$\\

0 & $k_{j6}=-(\frac{L_{j4}}{L_{j2}})^{2}$ & $a_{j6} = k_{j5} + k_{j6}$ & $b_{j6}=(1+z_{j1})k_{j5}+(3+z_{j0})k_{j6}+z_{j2}k_{j4}$\\

0 & 0 & $a_{j7}=  k_{j6}$ & $b_{j7}=(1+z_{j1})k_{j6}+(3+z_{j0})k_{j7}+z_{j2}k_{j5}$\\

0 & 0 & 0 & $b_{j8}= z_{j2}k_{j6}$\\
 \hline
\end{tabular}\\
\end{table}

\begin{table}
\begin{tabular}{ |p{9cm}| |p{2.5cm}| |p{5cm}| }
 \hline
 \multicolumn{3}{|c|}{Table 7: Coefficients for the eigenvector $V_{j}$, $j$ = 3 in Eq. (A.44)}\\
 \hline
$k_{jk}; j=3$ & $a_{jk}; j=3$ & $b_{jk}; j =3$\\
 \hline
0 & $a_{30}=1$ & $b_{30}= 3 + z_{30}$\\

$k_{31}=-\frac{L_{31}}{L_{30}}$ &$a_{31}= 1 + k_{31}$ & $b_{31}= (1+z_{31})+(3+z_{30})k_{31}$\\

$k_{32}=-\frac{L_{32}}{L_{30}}+(\frac{L_{31}}{L_{30}})^{2}$ & $a_{32}= k_{31} + k_{32}$ & $b_{32}=(1+z_{31})k_{31}+(3+z_{30})k_{32}+z_{32}$\\

$k_{33}=-\frac{L_{33}}{L_{30}}+\frac{2 L_{31} L_{32}}{L_{30}^{2}}-(\frac{L_{31}}{L_{30}})^{3}$ & $a_{33}= k_{32} + k_{33}$ & $b_{33}=(1+z_{31})k_{32}+(3+z_{30})k_{33}+z_{32}k_{31}$\\

$k_{34}=-\frac{L_{34}}{L_{30}}+(\frac{2 L_{31} L_{33}}{L_{30}^{2}}+(\frac{L_{32}}{L_{30}})^{2})-(\frac{3 L_{31}^{2} L_{32}}{L_{30}^{3}}+\frac{3 L_{31} L_{32}^{2}}{L_{30}^{3}})+(\frac{L_{31}}{L_{30}})^{4}$ & $a_{34}= k_{33} + k_{34}$ & $b_{34}=(1+z_{31})k_{33}+(3+z_{30})k_{34}+z_{32}k_{32}$\\

$k_{35}=(\frac{2 L_{31} L_{34}}{L_{30}^{2}}+\frac{2 L_{32} L_{33}}{L_{30}^{2}})-\frac{3 L_{31}^{2} L_{33}}{L_{30}^{3}}+(\frac{4 L_{31}^{3}L_{32}}{L_{30}^{4}}-(\frac{L_{31}}{L_{30}})^{5})$ & $a_{35}= k_{34} + k_{35}$ & $b_{35}=(1+z_{31})k_{34}+(3+z_{30})k_{35}+z_{32}k_{33}$\\
 \hline
\end{tabular}\\
\end{table}

\begin{table}
\begin{tabular}{ |p{9cm}||p{7cm}| }
 \hline
 \multicolumn{2}{|c|}{Table 8: Coefficients for $L_{j2}; j = 1, 2$ and the determinant of the eigenvector matrix $\bf{M}$}\\
 \hline
$L_{j2};\enspace j = 1,\enspace 2$ & det($\bf{M}$)\\
 \hline
$F_{j1}(p)=(4+2z_{j0}(p))f_{j1}(p)$ & $T_{1}(p)= p(-\frac{m_{1}}{F_{11}}-\frac{m_{5}}{F_{21}})$\\

$F_{j2}(p)=(4+2z_{j0}(p))f_{j2}(p)+(z_{j1}(p)^{2}+2z_{j1}(p)+(1-p^{2}))$ & $T_{2}(p)= p(-\frac{m_{2}}{F_{11}}+\frac{f_{11}}{F_{11}}\frac{1}{(2z_{10}+4)}-\frac{m_{6}}{F_{21}}-\frac{m_{11}}{L_{30}})$\\
 \hline
\end{tabular}
\end{table}

\begin{table}
\begin{tabular}{ |p{7.5cm}| |p{4.5cm}| |p{4.0cm}| }
 \hline
 \multicolumn{3}{|c|}{Table 9: Components $V_{ij}; i, j$ = 1, 2, 3 of the eigenvector matrix $\bf{M}$ in Eq. (A.71)}\\
 \hline
$V_{1i}$ & $V_{2i}$ & $V_{3i}$\\
 \hline
$V_{11} = p[-\frac{1}{F_{11}}+\frac{f_{11}}{F_{11}}\frac{1}{(2z_{10}+4)}\frac{1}{\bar{n}^{2}}(\frac{\Delta}{\gamma})^{2}]\bar{n}(\frac{\Delta}{\gamma})^{-1}$ & $V_{21} = -\frac{p}{F_{21}}\bar{n}(\frac{\Delta}{\gamma})^{-1}$ & $V_{31}= -\frac{p}{L_{30}} (\frac{1}{\bar{n}})(\frac{\Delta}{\gamma})$\\

$V_{12} = [\frac{(3+z_{10})}{F_{11}}+(\frac{f_{11}}{F_{11}})\frac{(1+z_{10})}{(4+2z_{10})}(\frac{1}{\bar{n}})^{2}(\frac{\Delta}{\gamma})^{2}]\bar{n}(\frac{\Delta}{\gamma})^{-1}$ & $V_{22}= \frac{(3+z_{20})}{F_{21}}\bar{n}(\frac{\Delta}{\gamma})^{-1}$ & $V_{32}= \frac{(3+z_{30})}{L_{30}}(\frac{1}{\bar{n}})(\frac{\Delta}{\gamma})$\\

$V_{13}= 1$ & $V_{23}= 1$ & $V_{33}= 1$\\
 \hline
\end{tabular}\\
\end{table}

\begin{table}
\begin{tabular}{ |p{5.5cm}| |p{5.0cm}| |p{6.0cm}| }
 \hline
 \multicolumn{3}{|c|}{Table 10: Cofactors $T_{ij}; i, j$ = 1, 2, 3 of the eigenvector matrix $\bf{M}$ in Eq. (A.71)}\\
 \hline
$T_{1i}$ & $T_{2i}$ & $T_{3i}$\\
 \hline
$T_{11} = [m_{1}(p)+m_{2}(p)\frac{1}{\bar{n}^{2}}(\frac{\Delta}{\gamma})^{2}]\bar{n}(\frac{\Delta}{\gamma})^{-1}$ & $T_{21} = [m_{5}(p)+m_{6}(p)\frac{1}{\bar{n}^{2}}(\frac{\Delta}{\gamma})^{2}]\bar{n}(\frac{\Delta}{\gamma})^{-1}$ & $T_{31}= [m_{11}(p)+m_{12}(p)\frac{1}{\bar{n}^{2}}(\frac{\Delta}{\gamma})^{2}]\bar{n}(\frac{\Delta}{\gamma})^{-1}$\\

$T_{12} = [m_{3}(p)+m_{4}(p)\frac{1}{\bar{n}^{2}}(\frac{\Delta}{\gamma})^{2}]\bar{n}(\frac{\Delta}{\gamma})^{-1}$ & $T_{22}= [m_{7}(p)+m_{8}(p)\frac{1}{\bar{n}^{2}}(\frac{\Delta}{\gamma})^{2}]\bar{n}(\frac{\Delta}{\gamma})^{-1}$ & $T_{32}= [m_{13}(p)+m_{14}(p)\frac{1}{\bar{n}^{2}}(\frac{\Delta}{\gamma})^{2}]\bar{n}(\frac{\Delta}{\gamma})^{-1}$\\

$T_{13}= \frac{p}{F_{21}L_{30}}(z_{20}-z_{30})$ & $T_{23}= [m_{9}(p)+m_{10}(p)\frac{1}{\bar{n}^{2}}(\frac{\Delta}{\gamma})^{2}]$ & $T_{33}= [m_{15}(p)+m_{16}(p)\frac{1}{\bar{n}^{2}}(\frac{\Delta}{\gamma})^{2}]\bar{n}^{2}(\frac{\Delta}{\gamma})^{-2}$\\
 \hline
\end{tabular}\\
\end{table}

\begin{table}
\begin{tabular}{ |p{3cm}| |p{4.5cm}| |p{4.5cm}| |p{4cm}|}
 \hline
 \multicolumn{4}{|c|}{Table 11: Coefficients $m_{i}(p)$ for the cofactors $T_{ij}$ of the eigenvector matrix $\bf{M}$}\\
 \hline
$m_{1}(p)=\frac{(3+z_{20})}{F21}$ & $m_{5}(p)=-\frac{(3+z_{10})}{F_{11}}$ & $m_{9}(p)=\frac{p}{F_{11}L_{30}}(z_{10}+z_{30}+6)$ & $m_{13}(p)=p(\frac{1}{F_{11}}-\frac{1}{F_{21}})$\\

$m_{2}(p)=-\frac{(3+z_{30})}{F31}$ & $m_{6}(p)=-(\frac{f_{11}}{F_{11}}\frac{(1+z_{10})}{(4+2z_{10})}-\frac{(3+z_{30})}{L_{30}})$ & $m_{10}(p)= p(\frac{f_{11}}{F_{11}L_{30}})\frac{(z_{10}-z_{30}-2)}{(2z_{10}+4)}$ & $m_{14}(p)= -(\frac{f_{11}}{F_{11}}\frac{1}{(2z_{10}+4)})$\\

$m_{3}(p)=\frac{p}{F21}$ & $m_{7}(p)=-\frac{p}{F_{11}}$ & $m_{11}(p)=(\frac{(3+z_{10})}{F_{11}}-\frac{(3+z_{20})}{F_{21}})$ & $m_{15}(p)=p(\frac{(z_{10-z_{20}})}{F_{11}F_{21}})$\\

$m_{4}(p)=-\frac{p}{L_{30}}$ & $m_{8}(p)= p (\frac{f_{11}}{F_{11}}\frac{1}{(2z_{10}+4)}+\frac{1}{L_{30}})$ & $m_{12}(p)= (\frac{f_{11}}{F_{11}})\frac{(1+z_{10})}{(4+2z_{10})}$ & $m_{16}(p)= p(\frac{f_{11}}{F_{11}F_{21}}\frac{(z_{10}+z_{20}+4)}{(2z_{10}+4)})$\\
 \hline
\end{tabular}\\
\end{table}

\begin{table}
\begin{tabular}{ |p{6.5cm}| |p{6.5cm}| |p{3cm}| }
 \hline
 \multicolumn{3}{|c|}{Table 12: Coefficients $A_{i}(p)$, $B_{i}(p)$ and $C_{i}(p)$ for the population and coherence terms in Eqs. (A.99) to (A.101) }\\
 \hline
$A_{i}$ & $B_{i}$ & $C_{i}$\\
 \hline
$A_{1}= -\frac{P}{F_{11}}(m_{1}+pm_{3})$ & $B_{1}= \frac{(3+z_{10})}{F_{11}}(m_{1}+pm_{3})$ & $C_{1}= (m_{1}+pm_{3})$\\

$A_{2}= p(\frac{f_{11}}{F_{11}}\frac{1}{(2z_{10}+4)}(m_{1}+pm_{3})-\frac{1}{F_{11}}(m_{2}+pm_{4}))$ & $B_{2}= (\frac{f_{11}}{F_{11}}\frac{1}{(2z_{10}+4)}(m_{1}+pm_{3})+\frac{(3+z_{10})}{F_{11}}(m_{2}+pm_{4}))$ & $C_{2}= (m_{2}+pm_{4})$\\

$A_{3}= -\frac{p}{F_{21}}(m_{5}+pm_{7})$ & $B_{3}= \frac{(3+z_{20})}{F_{21}}(m_{5}+pm_{7})$ & $C_{3}= (m_{5}+pm_{7})$\\

$A_{4}= -\frac{p}{F_{21}}(m_{6}+pm_{8})$ & $B_{4}= \frac{(3+z_{20})}{F_{21}}(m_{6}+pm_{8})$ & $C_{4}= (m_{6}+pm_{8})$\\

$A_{5}= -\frac{p}{L_{30}}(m_{11}+pm_{13})$ & $B_{5}= \frac{(3+z_{30})}{L_{30}}(m_{11}+pm_{13})$ & $C_{5}= (m_{11}+pm_{13})$\\

$A_{6}= -\frac{p}{L_{30}}(m_{12}+pm_{14})$ & $B_{6}= \frac{(3+z_{30})}{L_{30}}(m_{12}+pm_{14})$ & $C_{6}= (m_{12}+pm_{14})$\\
 \hline
\end{tabular}\\
\end{table}

 }



\clearpage
\newpage


\begin{thebibliography}{99}



\bibitem{Schlosshauer}
M. Schlosshauer, {\it Decoherence and  The Quantum-to-Classical Transition} (Springer-Verlag Berlin Heidelberg, 1997).

\bibitem{BPbook}
H.-P. Breuer and F. Petruccione, {\it The Theory of Open Quantum Systems} (Clarendon Press, Oxford, 2006), Chap. 3.4.

\bibitem{ScullyBook}
M. O. Scully and M. S. Zubairy, {\it Quantum Optics} (Cambridge University Press, Cambridge, UK, 1997).

\bibitem{Qsensors}
C. L. Degen, F. Reinhard, and P. Cappellaro, Quantum sensing, Rev. Mod. Phys. {\bf 89}, 035002 (2017).

\bibitem{Ladd}
T. D. Ladd, F. Jelezko, R. Laflamme, Y. Nakamura, C. Monroe, and J. L. O'Brien,  Quantum Computers, Nature (London) {\bf 464}, 45 (2010).

\bibitem{Scully92}
M. Fleischhauer, C. H. Keitel, M. O. Scully, and C. Su, Lasing without inversion and enhancement of the index of refraction via interference of incoherent pump processes, Opt. Commun {\bf 87}, 109 (1992). 

\bibitem{HegerfeldtPlenio93}
G. C. Hegerfeldt and M. B. Plenio, Coherence with incoherent light: A new type of quantum beats for a single atom, \pra{47}{2186}{1993}.

\bibitem{AgarwalMenon01}
G. S. Agarwal and S. Menon,  Quantum interferences and the question of thermodynamic equilibrium, \pra{63}{023818}{2001}.

\bibitem{Scully06}
V. V. Kozlov, Y. Rostovtsev, and M. O. Scully, Inducing quantum coherence via decays and incoherent pumping with application to population trapping, lasing without inversion, and quenching of spontaneous emission,  \pra{74}{063829}{2006}.

\bibitem{Ou08} 
B.-Q. Ou, L.-M. Liang, and C.-Z. Li, Coherence induced by incoherent pumping field and decay process in three-level $\Lambda$-type atomic system Opt. Commun. {\bf 281}, 4940 (2008).

\bibitem{Scully11}
M.~O.~Scully, K. R. Chapin, K.E.  Dorfman, M. B. Kim, and A. Svidzinsky, Quantum Heat Engine Power can be Increased by Noise-Induced Coherence, Proc. Natl. Acad. Sci. USA {\bf 108}, 15097 (2011).

\bibitem{Scully13}
K. E. Dorfman, D. V. Voronine, S. Mukamel, and M. O. Scully, Photosynthetic reaction center as a quantum heat engine, Proc. Natl. Acad. Sci. USA {\bf 110}, 2746 (2013).



\bibitem{prl14}
T. V. Tscherbul and P. Brumer, Long-lived Quasistationary Coherences in a $V$-type System Driven by Incoherent Light,  {Phys. Rev. Lett.} {\bf 113}, 113601 (2014). 

\bibitem{Amr16}
A. Dodin, T. V. Tscherbul, and P. Brumer, Quantum dynamics of incoherently driven V-type systems: Analytic solutions beyond the secular approximation, \jcp{144}{244108}{2016}.

\bibitem{Amr16slow}
A. Dodin, T. V. Tscherbul, and P. Brumer,
Coherent dynamics of V-type systems driven by time-dependent incoherent radiation, \jcp{145}{244313}{2016}.


\bibitem{Keitel05}
M. Macovei, J. Evers, and C. H. Keitel, Quantum correlations of an atomic ensemble via an incoherent bath, \pra{72}{063809}{2005}.

\bibitem{AgarwalTract}
G. S. Agarwal, {\it Quantum Statistical Theories of Spontaneous Emission and their Relation to Other Approaches} (Springer-Verlag, Berlin, 1974).

\bibitem{Evers13}
K. P. Heeg {\it et al.}, Vacuum-Assisted Generation and Control of Atomic Coherences at X-Ray Energies, \prl{111}{073601}{2013)}


\bibitem{Chin13}
C. Creatore, M. A. Parker, S. Emmott, and A. W. Chin, Efficient Biologically Inspired Photocell Enhanced by Delocalized Quantum States, \prl{111}{253601}{2013}.

\bibitem{Chin16}
A. Fruchtman, R. G\'omez-Bombarelli, B. W. Lovett, and E. M. Gauger, Photocell Optimization Using Dark State Protection, \prl{117}{203603}{2016}.

\bibitem{Blum}
K. Blum, {\it Density Matrix Theory and Applications}  (Springer, 2011).



\bibitem{AmrCa17}
A. Dodin,  T. V. Tscherbul, R. Alicki, A. Vutha, and P. Brumer, Non-secular Redfield Dynamics and Fano Coherences in Incoherent Excitation: An Experimental Proposal, arXiv:1711.10074v1  (2017).

\bibitem{BoyceDiPrima}
W. E. Boyce and R. C. DiPrima, {\it Elementary Differential Equations} (Wiley, NY, 2008).














\bibitem{RydbergBook}
T. F. Gallagher, {\it Rydberg Atoms} (Cambridge University Press, Cambridge, UK, 1994).

\bibitem{RydbergCMB}
T. V. Tscherbul and P. Brumer, Coherent dynamics of Rydberg atoms in cosmic-microwave-background radiation, \pra{89}{013423}{2014}.



\end{thebibliography}
\end{document}